\documentclass[fleqn,usenatbib]{mnras}

\usepackage{newtxtext,newtxmath}

\usepackage[T1]{fontenc}

\DeclareRobustCommand{\VAN}[3]{#2}
\let\VANthebibliography\thebibliography
\def\thebibliography{\DeclareRobustCommand{\VAN}[3]{##3}\VANthebibliography}

\usepackage{graphicx}	
\usepackage{amsmath}	
\usepackage{caption}

\title[Resolved ISM properties in NGC 3627]{Resolved ISM properties and scaling relations in the barred galaxy NGC~3627: constraints from NIKA2 observations}

\author[S.~Katsioli et al.]{S.~Katsioli$^{1,2}$\thanks{E-mail: s.katsioli@noa.gr},
E.~M.~Xilouris$^{1}$,
F.~Galliano$^{3}$,
R.~Adam$^{4}$,
P.~Ade$^{5}$,
H.~Ajeddig$^{3}$,
S.~Amarantidis$^{6}$,
P.~Andr\'e$^{3}$,
\newauthor H.~Aussel$^{3}$,
M.~Baes$^{7}$,
A.~Beelen$^{8}$,
A.~Beno\^it$^{9}$,
S.~Berta$^{10},$
A.~Bongiovanni$^{6}$,
J.~Bounmy$^{11},$
O.~Bourrion$^{11},$
\newauthor M.~Calvo$^{9}$,
A.~Catalano$^{11}$,
D.~Ch\'erouvrier$^{11}$,
I.~De~Looze$^{7}$,
M.~De~Petris$^{12}$,
F.-X.~D\'esert$^{13}$,
S.~Doyle$^{5}$,
\newauthor E.~F.~C.~Driessen$^{10}$,
G.~Ejlali$^{14}$,
A.~Ferragamo$^{12}$,
A.~Gomez$^{15}$, 
J.~Goupy$^{9}$,
C.~Hanser$^{11}$,
\newauthor A.~Hughes$^{16}$,
A.~P.~Jones$^{17}$,
F.~K\'eruzor\'e$^{18}$,
C.~Kramer$^{10}$,
B.~Ladjelate$^{6}$,
G.~Lagache$^{8}$,
S.~Leclercq$^{10}$,
\newauthor J.-F.~Lestrade$^{19}$,
J.~F.~Mac\'ias-P\'erez$^{11}$,
S.~C.~Madden$^{3}$,
A.~Maury$^{20,21}$,
F.~Mayet$^{11}$,
A.~Monfardini$^{9}$,
\newauthor A.~Moyer-Anin$^{11}$,
M.~Mu\~noz-Echeverr\'ia$^{16}$,
I.~Myserlis$^{6}$,
A.~Nersesian$^{7,22}$,
A.~Paliwal$^{23}$,
L.~Pantoni$^{7,24}$,
\newauthor D.~Paradis$^{16}$,
L.~Perotto$^{11}$,
G.~Pisano$^{12}$,
N.~Ponthieu$^{13}$,
V.~Rev\'eret$^{3}$,
A.~J.~Rigby$^{25}$,
A.~Ritacco$^{26,27}$,
\newauthor H.~Roussel$^{28}$,
F.~Ruppin$^{29}$,
M.~S\'anchez-Portal$^{6}$,
S.~Savorgnano$^{11}$,
K.~Schuster$^{10}$,
A.~Sievers$^{6}$,
\newauthor M.~W.~L.~Smith$^{5}$,
F.~Tabatabaei$^{14}$,
J.~Tedros$^{6}$,
C.~Tucker$^{5}$,
N.~Ysard$^{17}$,
and R.~Zylka$^{10}$
\\
(Affiliations can be found after the references)
}

\date{Accepted XXX. Received YYY; in original form ZZZ}

\pubyear{\the\year{}}

\begin{document}
\label{firstpage}
\pagerange{\pageref{firstpage}--\pageref{lastpage}}
\maketitle

\begin{abstract}
We investigate the interplay between star formation, interstellar medium (ISM) components, and dust properties in NGC~3627 using new NIKA2 1.15 and 2~mm observations from the IMEGIN Large Program. Our goal is to analyze dust and radio emission, decompose contributions in the millimeter-centimeter regime, and explore ISM properties within the galaxy. We perform spectral energy distribution fitting, at both global and spatial scales, using the \texttt{THEMIS} dust model within the \texttt{HerBIE} code, applied to data from 3.4~$\mu$m to 6~cm. We decompose emission into dust, free-free, and synchrotron components, and examine correlations with gas surface density and star formation activity. Additionally, we analyze the small dust grain fraction and its variation across the galaxy. We find $\sim$10\% radio emission at 2~mm, peaking at 18\% in the southern bar-end, which hosts the highest star formation activity. However, an isolated star-forming region beyond this bar-end is the most efficient, as indicated by its elevated dust production efficiency and effective yield, predicted by our simplistic dust evolution model. The 160~$\mu$m emission shows the strongest correlation with molecular gas, while 1.15~mm better traces the dust mass surface density. Small grains, which make up $\sim$13\% of dust mass (2~$\times$~10$^7$~M$_{\odot}$), are depleted in intense radiation fields, with a notable deficit in the southern tidal tail. ISM properties and chemical evolution indicate that dynamical processes, such as bar-driven gas flows and tidal interactions, are crucial in shaping the galactic structure, influencing star formation efficiency, and dust distribution.

\end{abstract}




\section{Introduction}\label{sec:intro}

Thermal emission from dust, and radio continuum emission, as well as gas line transitions in galaxies, are key tracers of the physical mechanisms that are taking place inside galaxies and help us understand the complex processes that are shaping the interstellar medium (ISM). The diverse wavelength ranges of the Spectral Energy Distribution (SED) of galaxies are associated with properties of specific components of the ISM or a combination of them. One of the most complex spectral regions is the millimeter (mm) domain, as it encompasses contributions from multiple ISM components, including thermal dust emission, synchrotron radiation, free-free emission, and molecular lines \cite[see, e.g.,][]{2023A&A...679A...7K}.

Despite the fact that cold dust and radio emission (mainly the combination of synchrotron and free-free mechanisms) are the main sources of mm-wavelength radiation in galaxies, other, more exotic, features are present in this wavelength range which require deeper understanding. Excess emission in the sub-millimeter (submm) to mm wavelengths, relative to contemporary, state-of-the-art models, has extensively been reported in the past \citep[e.g.,][]{2003A&A...407..159G, 2005A&A...434..867G, 2018ARA&A..56..673G, 2005ApJ...633..272B, 2009ApJ...706..941Z, 2011A&A...534A.118P, 2011A&A...535A.124C, 2012MNRAS.425..763G, 2013A&A...557A..95R, 2016A&A...590A..56H, 2022A&A...666A.192D, 2023A&A...679A...7K} with the likely explanation being modification of the dust properties, thus, model dependent. 
Anomalous Microwave Emission (AME), on the other hand, is an excess of microwave emission observed at frequencies between $\sim 10$ and 60~GHz, which cannot be fully explained by standard emission mechanisms and is thought to originate from spinning dust grains \citep[very small dust particles, the current likely hypothesis being polycyclic aromatic hydrocarbons (PAHs), that rotate rapidly and emit electric dipole radiation, e.g.,][]{2014A&A...565A.103P, 2016A&A...594A..25P, 2018NewAR..80....1D, 2021MNRAS.503.2927C, 2022A&A...658L...8B}.

\begin{table}
\begin{center}
\captionsetup{labelfont=bf} 
\caption{Morphological and physical properties of NGC~3627.} 
\label{tab:prop}
\begin{tabular}{llc}
\hline \hline
 \textbf{Properties}   & \textbf{Values} & \textbf{References} \\ \hline
 RA$_\mathrm{J2000}$ & 11$^\mathrm{h}$20$^\mathrm{m}$14.96$^\mathrm{s}$ & (a) \\
 DEC$_\mathrm{J2000}$ & +12$^{\circ}$59$^{\prime}$29.5$^{\prime\prime}$ & (a) \\
 Hubble type & SAB(s)b & (b) \\
 Nuclear type & LINER/Seyfert 2 & (c) \\
 Distance~[Mpc] & 11.32~$\pm$~0.48 & (d) \\
 Linear scale [pc/$^{\prime\prime}$] & 54.9~$\pm$~2.3 & (d) \\
 Inclination & 57.3$^{\circ}$ & (a) \\
 Positional angle & 173$^{\circ}$ & (e) \\
 log$_{10}(M_{\star})$~[M$_{\odot}$] & 10.84 & (f) \\
 log$_{10}(M_\mathrm{HI})$~[M$_{\odot}$] & 9.04 & (g) \\
 log$_{10}(M_\mathrm{H_2})$~[M$_{\odot}$] & 9.65 & (g) \\
 log$_{10}$(SFR)~[M$_{\odot}$~yr$^{-1}$] & 0.59 & (f) \\
    \\
\hline
\end{tabular}
\end{center}

\begin{minipage}{\linewidth}
\small
\textbf{References.} (a) \cite{2022ApJS..258...10L}; (b) \cite{1991rc3..book.....D}; (c) \cite{2022ApJ...936..162S}; (d) \cite{2021MNRAS.501.3621A}; (e) \cite{2020ApJ...897..122L}; (f) \cite{2021ApJS..257...43L} using GALEX UV and WISE IR photometry; (g) this study.
\end{minipage}
\end{table}

Nearby galaxies serve as ideal laboratories for studying these processes, as their diverse structures — nucleus, stellar bar, spiral arms, and halo — offer distinct local dynamical and physical environments for such analysis.
NGC~3627 is a well-observed barred spiral galaxy \citep[SAB(s)b;][]{1991rc3..book.....D} whose proximity \citep[$d$~=~11.32~Mpc;][]{2021MNRAS.501.3621A} and large angular size \citep[$R_{25}\sim5.1^{\prime}$;][]{2017MNRAS.466...49J} make it an excellent candidate for studying the ISM components in various galactic environments. NGC~3627 is a massive star-forming galaxy at an inclination of 57.3$^{\circ}$ \citep{2022ApJS..258...10L} which likely harbors a low-luminosity Seyfert~2 Active Galactic Nucleus \citep[AGN; e.g.,][]{1997ApJS..112..315H, 1998AAS...193.0607P, 2022ApJ...936..162S, 2023MNRAS.526.6347D}. However, it has also been classified as a Low-Ionization Nuclear Emission-line Region \citep[LINER; e.g.,][]{2022ApJ...936..162S}. The main properties of NGC~3627 are summarized in Table~\ref{tab:prop}.

The first attempt to map NGC~3627 at mm wavelengths was conducted by \cite{1994A&A...281..681S} using the 7-channel MPfIR bolometer on the IRAM 30-m telescope. Their 1.28~mm mapping revealed cold dust emission, primarily concentrated in the central bar and the spiral arms, and well aligned with the overall CO~(1-0) and CO~(2-1) line emission morphology observed with the IRAM 30-m telescope \citep{1996A&A...306..721R} and ALMA \citep{2021ApJS..257...43L}, respectively.
\cite{2021MNRAS.506..963B} also showed, using NOEMA observations of CO isotopomers and dense gas tracers at resolutions of $\sim$100~pc, that the molecular gas properties in NGC~3627 vary significantly with the environment, highlighting the complexity of its ISM.
More recently its mm emission has been investigated in high resolution, in specific regions through observations at 3~mm and 9~mm with ALMA and the VLA, respectively \citep{2015ApJ...813..118M}, and at 3.3~mm with the GBT/MUSTANG camera \citep{2012MNRAS.425.1257N}.
By constraining the free-free emission in various regions, \citet{2012MNRAS.425.1257N} derived a typical free-free fraction of $80\%~\pm~20\%$ at 3.3~mm. They estimated the star formation rates (SFRs), and found that the southern bar-end is the most active $[(0.64~\pm~0.14)$~M$_{\odot}$yr$^{-1}]$, followed by the northern bar-end $[(0.58\pm0.19)$~M$_{\odot}$~yr$^{-1}]$.
According to \cite{2015ApJ...813..118M}, free-free emission dominates at 9~mm, accounting for an average of (79~$\pm$~19)\% when evaluating both the nucleus and two observed star-forming regions in NGC~3627. This study further suggested that the galaxy's isolated star-forming region in the southern part has a higher efficiency in converting molecular gas into stars.

Thermal dust emission in NGC~3627 has also been studied via SED fitting \citep[e.g.,][]{2012MNRAS.425..763G, 2014MNRAS.439.2542G, 2015A&A...582A.121R, 2020ApJ...889..150A, 2020MNRAS.496.3668D}. \cite{2012MNRAS.425..763G} showed that areas where cold dust temperatures reach maximum values coincide with the 24~$\mu$m emission peaks, suggesting that star-forming regions contribute more significantly to heating the cold dust compared to the interstellar radiation field (ISRF). In their study, \cite{2014MNRAS.439.2542G} examined the presence of submm excess at 870~$\mu$m (by modeling the emission with two modified blackbodies) and found that in the case of NGC~3627, an excess, above the model predictions, is observed at this wavelength which can be attributed to cold dust emission combined with emission from the CO~(3-2) line.

The galaxy's synchrotron polarization morphology has been mapped \citep[e.g.,][]{2001A&A...378...40S}, and SED fitting at radio wavelengths has provided insights into its synchrotron emission distribution \citep[e.g.,][]{2017ApJ...836..185T}. An interesting discontinuity in the gas flow and the magnetic field direction in the eastern spiral arm of the galaxy has been reported \citep{2001A&A...378...40S}. \cite{2023MNRAS.519.1068L} found a strong correlation between synchrotron polarization and CO~(2-1) polarization, suggesting either a short diffusion length for cosmic ray electrons or that synchrotron emission is primarily influenced by strong magnetic fields in star-forming regions. On the other hand, NGC~3627 is one of a sample of galaxies that have been studied for AME after making use of K-band observations at 1.2 and 1.6~cm obtained with the Sardinia Radio Telescope \citep{2022A&A...658L...8B}. Although in this study, only upper limits were calculated for AME, the emissivity was found to be consistent with the average AME emissivity from a few previous detections. 

The asymmetries observed in its spiral arms indicate that NGC~3627 has undergone past interactions.
\cite{1978AJ.....83..219R} proposed that a tidal encounter with NGC~3628, a member of the Leo Triplet, occurred approximately 800~Myr ago \citep[see also][]{1979ApJ...229...83H, 2001A&A...378...40S, 2022A&A...658A..25W}.
More recently, \cite{2012A&A...544A.113W} analyzed X-ray data alongside the magnetic field structures and proposed that NGC~3627 experienced a more recent interaction with a dwarf galaxy, possibly along the eastern spiral arm, a few tens of Myr ago \citep[see also,][]{1993ApJ...418..100Z, 2003A&A...405...89C, 2011ASPC..446..111D}.

    \begin{figure*}
    \centering
    \captionsetup{labelfont=bf}
    \includegraphics[width=0.9\textwidth]{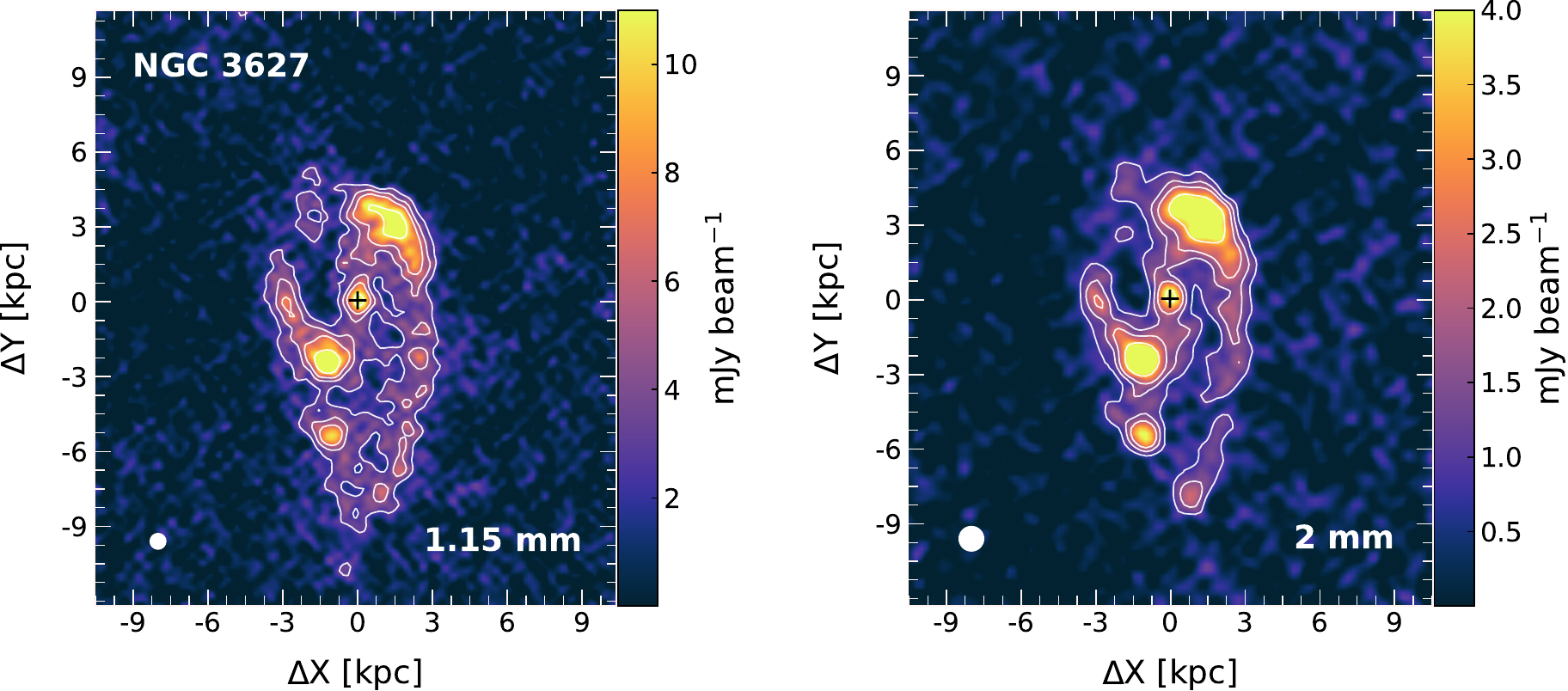}
    \caption{NGC~3627 mapped with the NIKA2 camera at 1.15~mm (left panel) and 2~mm (right panel) and with native resolutions of 11.1$^{\prime\prime}$ and 17.6$^{\prime\prime}$ ($\sim0.6$~kpc and $\sim1$~kpc, respectively; see the white circles in the bottom-left corner in each panel). The maps are centered at RA$_\mathrm{J2000}$~$=11^\mathrm{h}20^\mathrm{m}15^\mathrm{s}$, DEC$_\mathrm{J2000}$~$=+12^{\circ}59^{\prime}30^{\prime\prime}$. The surface brightness contours correspond to 3, 5, 7 and 12~$\times$~RMS. The RMS values are 0.9~mJy~beam$^{-1}$ and 0.3~mJy~beam$^{-1}$ at 1.15~mm and 2~mm respectively. The maps are presented in projected galactic distances $\Delta$X, $\Delta$Y from the center (marked by crosses), in kpc.
    }
    \label{fig:obs}
    \end{figure*}

\begin{table*}
\centering
\captionsetup{labelfont=bf}
\caption{Properties of the maps and integrated photometry used in the SED fitting of NGC~3627. }
\label{tab:photometry}
\begin{tabular}{lcccccc}
\hline \hline
{\bf Telescope /}  & {\bf Central } & {\bf Resolution} & {\bf Pixel} & {\bf $\pmb{\nu L_{\nu,\mathrm{Glob}}~\pm}$~RMS} & \textbf{$\pmb{\pm}$ Calibration}  & {\bf Reference} \\ 
{\bf Instrument}  & {\bf wavelength ($\pmb{\mu}$m)} & {\bf FWHM} & {\bf size} & \bf{(L$\pmb{_{\odot}}$)} & \bf{uncertainty} &  \\ \hline
 WISE \textsuperscript{(a)} & 3.4 & 6.1$^{\prime\prime}$ & 1.4$^{\prime\prime}$ & (6.63 $\pm$ 1.91) $\times 10^{9}$ & 3.2\% \textsuperscript{(i)} & This study \\
\emph{Spitzer} / IRAC \textsuperscript{(a)} & 3.6 & 1.7$^{\prime\prime}$ & 0.6$^{\prime\prime}$ & (6.43 $\pm$ 1.74) $\times 10^{9}$ & 10.2\% \textsuperscript{(i)} & This study \\
\emph{Spitzer} / IRAC \textsuperscript{(a)} & 4.5 & 1.7$^{\prime\prime}$ & 0.6$^{\prime\prime}$ & (3.30 $\pm$ 0.83) $\times 10^{9}$ & 10.2\% \textsuperscript{(i)} & This study \\
WISE \textsuperscript{(a)} & 4.6 & 6.4$^{\prime\prime}$ & 1.4$^{\prime\prime}$ & (3.02 $\pm$ 0.87) $\times 10^{9}$ & 3.5\% \textsuperscript{(i)} & This study \\
\emph{Spitzer} / IRAC \textsuperscript{(a)} & 5.8 & 1.9$^{\prime\prime}$ & 0.6$^{\prime\prime}$ & (5.91 $\pm$ 1.35) $\times 10^{9}$ & 10.2\% \textsuperscript{(i)} & This study \\
\emph{Spitzer} / IRAC \textsuperscript{(a)} & 8.0 & 2.0$^{\prime\prime}$ & 0.6$^{\prime\prime}$ & (1.06 $\pm$ 0.22) $\times 10^{10}$ & 10.2\% \textsuperscript{(i)} & This study \\
WISE \textsuperscript{(a)} & 11.6 & 6.5$^{\prime\prime}$ & 1.4$^{\prime\prime}$ & (4.70 $\pm$ 1.33) $\times 10^{9}$ & 5.0\% \textsuperscript{(i)} & This study \\
WISE \textsuperscript{(a)} & 22.1 & 12$^{\prime\prime}$ & 1.4$^{\prime\prime}$ & (2.48 $\pm$ 1.22) $\times 10^{9}$ & 7.0\% \textsuperscript{(i)} & This study \\
\emph{Spitzer} / MIPS \textsuperscript{(a)} & 24.0  & 6$^{\prime\prime}$ & 1.5$^{\prime\prime}$ & (3.62 $\pm$ 0.65) $\times 10^{9}$ & 4.0\% \textsuperscript{(i)} & This study \\
\emph{Herschel} / PACS \textsuperscript{(a)} & 70.0 & 6$^{\prime\prime}$ & 2$^{\prime\prime}$ & (1.87 $\pm$ 0.11) $\times 10^{10}$ & 5.4\% \textsuperscript{(i)} & This study \\
\emph{Herschel} / PACS \textsuperscript{(a)} & 100.0 & 8$^{\prime\prime}$ & 3$^{\prime\prime}$ & (2.29 $\pm$ 0.01) $\times 10^{10}$ & 5.4\% \textsuperscript{(i)} & This study \\
\emph{Herschel} / PACS \textsuperscript{(a)} & 160.0 & 12$^{\prime\prime}$ & 4$^{\prime\prime}$ & (1.50 $\pm$ 0.00) $\times 10^{10}$ & 5.4\% \textsuperscript{(i)} & This study \\
\emph{Herschel} / SPIRE \textsuperscript{(a)} & 250.0 & 18$^{\prime\prime}$ & 6$^{\prime\prime}$ & (4.40 $\pm$ 0.64) $\times 10^{9}$ & 5.9\% \textsuperscript{(i)} & This study \\
\emph{Herschel} / SPIRE \textsuperscript{(a)} & 350.0 & 25$^{\prime\prime}$ & 8$^{\prime\prime}$ & (1.27 $\pm$ 0.17) $\times 10^{9}$ & 4.3\% \textsuperscript{(i)} & This study \\
\emph{Planck} / HFI & 350.0 & & & (1.34 $\pm$ 0.01) $\times 10^{9}$ & 5.9\% \textsuperscript{(i)} & (1) \\
\emph{Herschel} / SPIRE \textsuperscript{(a)} & 500.0 & & & (2.90 $\pm$ 0.03) $\times 10^{8}$ & 5.9\% \textsuperscript{(i)} & This study \\
\emph{Planck} / HFI & 550.0 & & & (2.07 $\pm$ 0.03) $\times 10^{8}$ & 4.2\% \textsuperscript{(i)} & (1) \\
\emph{Planck} / HFI & 850.0 & & & (3.21 $\pm$ 0.18) $\times 10^{7}$ & 0.9\% \textsuperscript{(i)} & (1) \\
JCMT / SCUBA & 850.0 & & &(2.62 $\pm$ 0.98) $\times 10^{7}$ & 10.0\% \textsuperscript{(ii)} & (2) \\
\textbf{IRAM 30m / NIKA2} \textsuperscript{(b)} & \textbf{1150.0} & {\bf 11.1$\pmb{^{\prime\prime}}$} & {\bf 3$\pmb{^{\prime\prime}}$} & \textbf{(7.36 $\pmb{\pm}$ 0.54) $\pmb{\times 10^{6}}$} & \textbf{8.0\%  \textsuperscript{(iii)}} & \textbf{This study} \\
IRAS & 1300.0 & & & (6.44 $\pm$ 0.92) $\times 10^{6}$ & 10.0\% \textsuperscript{(ii)} & (3) \\
\emph{Planck} / HFI & 1380.0 & & & (5.00 $\pm$ 1.52) $\times 10^{6}$ & 0.5\% \textsuperscript{(i)} & (1) \\
\textbf{IRAM 30m / NIKA2} \textsuperscript{(b)} & \textbf{2000.0} & {\bf 17.6$\pmb{^{\prime\prime}}$} & {\bf 4$\pmb{^{\prime\prime}}$} & \textbf{(7.56 $\pmb{\pm}$ 1.37) $\pmb{\times 10^{5}}$} & \textbf{6.1\%  \textsuperscript{(iii)}} & \textbf{This study} \\
SRT & 12186.7 & & & (8.53 $\pm$ 1.28) $\times 10^{4}$ & 10.0\%  \textsuperscript{(ii)} & (4) \\
SRT & 16117.9 & & & (5.27 $\pm$ 1.04) $\times 10^{4}$ & 10.0\%  \textsuperscript{(ii)} & (4) \\
OVRO & 28018.0 & & & (4.27 $\pm$ 0.43) $\times 10^{4}$ & 10.0\%  \textsuperscript{(ii)} & (5) \\
\emph{Effelsberg} 100m & 28416.3 & & & (4.33 $\pm$ 0.42) $\times 10^{4}$ & 10.0\%  \textsuperscript{(ii)} & (6) \\
VLA \textsuperscript{(c)} & 35436.0 & 11$^{\prime\prime}$ & 1$^{\prime\prime}$ & (3.96 $\pm$ 0.12) $\times 10^{4}$ & 3.0\%  \textsuperscript{(iv)} & This study \\
\emph{Parkes} & 59958.5 & & & (3.73 $\pm$ 0.38) $\times 10^{4}$ & 10.0\%  \textsuperscript{(ii)} & (7) \\
GBO & 59958.5 & & & (3.49 $\pm$ 0.40) $\times 10^{4}$ & 10.0\%  \textsuperscript{(ii)} & (8)\\
VLA \textsuperscript{(c)} & 61812.9 & 13.5$^{\prime\prime}$ & 1$^{\prime\prime}$ & (2.73 $\pm$ 0.39) $\times 10^{4}$ & 3.0\%  \textsuperscript{(iv)} & This study \\
GBO & 61812.9 & & & (3.52 $\pm$ 0.11) $\times 10^{4}$ & 10.0\%  \textsuperscript{(ii)} & (9) \\
\emph{Effelsberg} 100m & 62456.8 & &  & (3.46 $\pm$ 0.78) $\times 10^{4}$ & 10.0\%  \textsuperscript{(ii)} & (10) \\
\hline
GALEX / FUV \textsuperscript{(a)} & 0.15  & 4.3$^{\prime\prime}$ & 3.2$^{\prime\prime}$ & & &\\
\textit{Chandra} / ACIS-S \textsuperscript{(d)} & 0.3 - 10 keV & 3.9$^{\prime\prime}$ & 3.9$^{\prime\prime}$ & & &  \\
BIMA / CO~(1-0) \textsuperscript{(e)} & 2600 & 6.1$^{\prime\prime}$ & 1$^{\prime\prime}$ & & & \\
IRAM / CO~(2-1) \textsuperscript{(f)} & 1300 & 13.4$^{\prime\prime}$ & 2$^{\prime\prime}$ & & & \\
VLA / \ion{H}{I} \textsuperscript{(g)} & 21 cm & 6$^{\prime\prime}$ & 1.5$^{\prime\prime}$ & & & \\
\hline
\end{tabular}
\begin{minipage}{\linewidth}
\small
\textbf{Notes.} The first part of the table lists the maps used in the resolved SED analysis (the ones with resolution and pixel scale information) with the remaining measurements used in the global SED fitting. The NIKA2 observations, presented here for the first time, are highlighted in bold. In the second part of the table, ancillary maps used in the analysis are reported.  Additionally, the table reports integrated photometry used in the global SED fitting, together with the corresponding 1$\sigma$ RMS uncertainties. 
The photometry values include measurements derived from the spatial maps as well as additional integrated fluxes from the literature, covering a wavelength range from 3.4~$\mu$m to 6.2~cm.  The data reduction process and the photometry method is described in detail in Sect.~\ref{sec:process}.
Sources of the spatial maps:
\textsuperscript{(a)} DustPedia \citep{2017PASP..129d4102D, 2018A&A...609A..37C};
\textsuperscript{(b)} This work;
\textsuperscript{(c)} \cite{2001A&A...378...40S};
\textsuperscript{(d)} \textit{Chandra} data archive;
\textsuperscript{(e)} \cite{2003ApJS..145..259H};
\textsuperscript{(f)} \cite{2009AJ....137.4670L};
\textsuperscript{(g)} \cite{2008AJ....136.2563W}.
Sources of the integrated photometry:
(1) DustPedia;
(2) \cite{2005ApJ...633..857D};
(3) \cite{1986A&A...157L...1C};
(4) \cite{2022A&A...658L...8B};
(5) \cite{1995A&AS..114...21N};
(6) \cite{1999A&A...345..461S};
(7) \cite{1970ApL.....5...29W};
(8) \cite{1975AJ.....80..771S};
(9) \cite{1991ApJS...75.1011G};
(10) \cite{2009ApJ...693.1392S}.
Sources of calibration uncertainties:
\textsuperscript{(i)} \cite{2021A&A...649A..18G};
\textsuperscript{(ii)} Assumed;
\textsuperscript{(iii)} \href{https://publicwiki.iram.es/PIIC/}{https://publicwiki.iram.es/PIIC/};
\textsuperscript{(iv)} \cite{2001A&A...369..380F}.
\end{minipage}
\end{table*}

Despite extensive studies of NGC~3627, key uncertainties remain regarding its mm emission, the processes driving its ISM, and the mechanisms and properties of its dust heating. 
These uncertainties highlight its complexity and make it a compelling target for further investigation. 
In the framework of the Interpreting the Millimetre Emission of Galaxies with IRAM and NIKA2 Large Program \citep[IMEGIN; PI: S. Madden; see][]{2023A&A...679A...7K, 2025A&A...693A..88E, pantoniprep} this study presents new observations of NGC~3627 at 1.15~mm and 2~mm.
The mm observations along with the auxiliary data used in this study are presented in Sect.~\ref{sec:data}. 
The steps of the data processing are described in Sect.~\ref{sec:process} and the multi-wavelength SED analysis of the galaxy is found in Sect.~\ref{sec:fitting}.
Using the SED modeled spectral luminosities from mm to cm wavelengths, we decompose the total emission into thermal dust, free-free, and synchrotron radiation in Sect.~\ref{sec:decomposition}.
In Sect.~\ref{sec:ISM}, the ISM properties in different regions throughout NGC~3627 are investigated, providing a quantitative analysis of all components and various scaling relations.
Finally, utilizing the state-of-the-art \texttt{THEMIS} dust model, in Sect.~\ref{subsec:dust_grains} we investigate the interstellar dust composition of small and large grains.
Our conclusions are summarized in Sect.~\ref{sec:conclu}.

\section{Data} \label{sec:data}

This study is part of the IMEGIN Large Program, which has targeted 22 nearby galaxies, at distances smaller than 25~Mpc, to map their emission at mm wavelengths. 
We used the New IRAM Kid Arrays~2 (NIKA2) camera installed on the \textit{Institut de Radio Astronomie Millim\'etrique} (IRAM) 30-m telescope (located at Pico Veleta in the Spanish Sierra Nevada at an altitude of 2850~m) to map their emission in two bands centered around 1.15 and 2~mm. High-resolution mapping of the mm continuum emission
enables a spatially resolved analysis of galaxies and their
ISM properties.
In the current study, we made use of the new NIKA2 observations of the IMEGIN star-forming NGC~3627, along with multi-wavelength archival data [from mid-infrared (MIR) to radio], obtained from both space and terrestrial facilities.

\subsection{NIKA2 observations and calibration} \label{subsec:obs}

NIKA2 \citep{adam2018, calvo2016, bourrion2016} is a dual-band camera mapping, simultaneously, at two frequency channels (150 and 260~GHz).
At both wavelengths, the NIKA2 camera maps the continuum intensity within broad transmission characterized of a $\sim50$~GHz full width at half maximum (FWHM).
It combines high sub-arcminute angular resolution [with half power beam widths (HPBWs) of 11.1$^{\prime\prime}$ and 17.6$^{\prime\prime}$ at 1.15~mm and 2~mm, respectively (corresponding to 0.6~kpc at 1.15~mm and 1~kpc at 2~mm at the distance of the galaxy)] and a wide field of view (FoV) of a $6.5^{\prime}$ diameter. NIKA2 consists of thousands of superconducting kinetic inductance detectors (KIDs) with two arrays of 1140 KIDs mapping the 1.15~mm emission and one array of 616 KIDs used for the 2~mm emission. The KIDs are cooled down to $\sim$150~mK.
NIKA2 was installed at the IRAM 30-m telescope in October 2015. The final commissioning campaign before the recent telescope upgrade was completed in April 2017 \citep[see][for the instrument calibration and performance]{2020A&A...637A..71P}.

NGC~3627 was observed on 18, 19 and 22 of February 2021, and 16, 17, 18 and 21 of March 2021, in the UT range between 20:00 and 04:00. A total of 12 hours were needed in order to conduct 34 on-the-fly (OTF) scans. As shown in \cite{2020A&A...637A..71P}, the time range of the observations was optimal in order to assure stable conditions. 
Weather conditions during our observations were stable with zenith opacities at 225~GHz [$\tau$(225~GHz)] ranging from 0.18 to 0.24 with a mean of 0.20 resulting in a mean precipitable water vapor (pwv) of the atmosphere of 3.38~mm.
The mapping strategy included consecutive scans at angles of $45^{\circ}$ and $-45^{\circ}$, relative to the major axis, to eliminate the presence of strip artifacts in the maps. The central detectors of each array covered an area of $22.2^{\prime}\times12.0^{\prime}$. The scanning was performed with a speed of 60$^{\prime\prime}$s$^{-1}$ and a spacing of 20$^{\prime\prime}$ between sub-scans.
Pointing and focus scans were conducted on nearby quasars about every one or two hours, as well as long scans obtained on primary and secondary calibrators during the pool observing campaign \citep[see][for a detailed description of the calibration method]{2020A&A...637A..71P}. The sky opacities have been measured every 5 minutes using an on-site taumeter operating at 225~GHz.

Using the {\tt piic/gildas}\footnote{\href{https://publicwiki.iram.es/PIIC/} {\label{piic} https://publicwiki.iram.es/PIIC/}} \citep{2013ascl.soft03011Z,berta-zylka2022} software (version 3/2021), the observations were combined and calibrated using the latest version of the calibration database and the appropriate data-associated files (DAFs). Instability corrections were applied to each OTF scan using a third-order polynomial baseline. 
A total of 40 iterations were performed, assuming pixels with a signal-to-noise ratio above two as sources during the iterations. 
Data beyond a radius of 350$^{\prime\prime}$ were neglected, as they suffer from low signal-to-noise ratios.  
No smoothing was applied during the source detection process to avoid artificial enhancement of low signal-to-noise regions.
Finally, interpolated opacity measurements were incorporated to correct for atmospheric absorption at the observed frequencies.

To estimate the total absolute flux uncertainty in the NIKA2 measurements of NGC~3627, a 5\% absolute flux uncertainty for the primary calibrators like Uranus was considered \citep[refer to][and references therein]{2020A&A...637A..71P}. This was then combined in quadrature with the relative RMS scatter observed in the calibrators as reduced using {\tt piic}.
During the two weeks of pool observations, the average RMS for both day and night, weighted by the weekly scan count, was found to be 6.3\% at 1.15~mm and 3.5\% at 2~mm. Consequently, the absolute flux uncertainty totaled 8.0\% at 1.15~mm and 6.1\% at 2~mm (refer to Table~\ref{tab:photometry}).

Fig.~\ref{fig:obs} shows the final produced maps at 1.15~mm and 2~mm, projected onto a grid of pixel sizes of 3$^{\prime\prime}$ and 4$^{\prime\prime}$, respectively. Galactic emission at both wavelengths is extended from the galaxy center up to 5~kpc in the north, 9~kpc in the south, and $3-4$~kpc in the east and west within the 3~$\times$~RMS level. The RMS values of the final maps are 0.9~mJy~beam$^{-1}$ and 0.3~mJy~beam$^{-1}$ at 1.15~mm and 2~mm, respectively. The morphology of the mm galactic emission in the two bands is comparable.
The nucleus appears luminous and is centrally located alongside a bar, with two noticeable spiral arms originating distinctly from the bar-ends, which are very bright at these wavelengths.
The galaxy at both 1.15~mm and 2~mm exhibits four obvious bright regions: the nucleus, the two bar-ends, and a large, isolated, \ion{H}{II} region \citep[see, e.g.,][]{2003A&A...405...89C}, $\sim$ 3~kpc to the south of the southern galactic bar-end. These four bright regions encompass 41\% and 51\% of the total emission of the galaxy at 1.15~mm and 2~mm, respectively. More specifically, 20\% and 10\% of the 1.15~mm emission originate from the northern and southern bar-ends, respectively, 8\% from the nucleus, and 7\% from the isolated \ion{H}{II} region, with the rest of the emission originating, primarily, from the spiral arms.
At 2~mm these fractions are 23\% and 12\% for the northern and southern bar-ends, respectively, 8\% for the nucleus, and 6\% for the large \ion{H}{II} region. The different emission components which contribute to the two wavelengths are thoroughly discussed in Sect.~\ref{sec:decomposition}.

Besides the difference of the two brightest regions (northern and southern bar-ends) in terms of integrated intensity, these two regions do not show the same morphology. The emitting region corresponding to the southern bar-end is smaller, while the northern one is larger and more distorted towards the spiral arm. 
An asymmetry is also evident between the two spiral arms of the galaxy, with the western spiral arm extending up to $\sim$12~kpc to the south, and the eastern spiral arm extending only up to $\sim$4.5~kpc to the north, with a solitary sign of its tail at distances of $\sim$6~kpc.

The overall morphology at the two bands is similar to the morphology of the galactic emission at the far-infrared/submillimeter (FIR/submm) spectral range (see Fig.~\ref{fig:maps}), and resembles the morphology of the CO~(1-0) line emission, which traces the H$_2$ distribution.
As indicated above, NGC~3627 has also been mapped at 1.28~mm by \cite{1994A&A...281..681S}, using the MPfIR 7-channel bolometer on the IRAM 30-m telescope, with the observed map in good agreement with the NIKA2 observation at 1.15~mm, presented in the current study. Although these observations do not go as deep as the NIKA2 observations, the general features discussed above on the differences between the two bar-ends seem to hold, with the northern one being about two times brighter than the southern one. 
The total photometry measured by \cite{1994A&A...281..681S} implies a spectral luminosity of $\nu L_\nu~=~(5.32~\pm~0.84)~\times~10^6$~L$_\odot$, consistent with the integrated SED derived in this study (see Sect.~\ref{sec:fitting}).

\subsection{Ancillary data} \label{subsec:archival_data}

We performed a multi-wavelength analysis of the galaxy by combining the new NIKA2 observations with archival data ranging, in wavelength, from 3.4~$\mu$m to 6.2~cm.
This wide range of data allows us to constrain the physical parameters associated with the thermal dust and radio emission by making use of the \texttt{HerBIE} SED fitting code (see Sect.~\ref{subsec:herbie}). We utilized the available data so that the SED modeling can be applied to both global and local scales within the galaxy.

In order to constrain the global SED of the galaxy, we measured the integrated flux for all the bands with available maps within a common elliptical aperture (see Sect.~\ref{sec:process}). For the rest bands, the photometrical data were retrieved from the literature and are presented in Table~\ref{tab:photometry}. The archival photometrical data used in the current analysis come from observations acquired from the \textit{Planck Space Telescope}, the \textit{James Clerk Maxwell Telescope} (JCMT), the VLA, the \textit{Effelsberg 100-m Radio Telescope}, the CSIRO \textit{Parkes Observatory}, the \textit{Green Bank Observatory} (GBO), the \textit{Owens Valley Radio Observatory} (OVRO), the \textit{Sardinia Radio Telescope} (SRT) and the \textit{Infrared Astronomical Satellite} (IRAS). 
The corresponding calibration uncertainties are also reported in Table~\ref{tab:photometry}. In the absence of reported values in the literature, we assumed a conservative calibration error of 10\%.

To fit the spatially resolved galactic emission, we compiled archival maps (from MIR to radio) obtained with the \textit{Spitzer Space Telescope} (SST), the \textit{Wide-field Infrared Survey Explorer} (WISE), the \textit{Herschel} and the \textit{Very Large Array} (VLA). The maps are presented in Fig.~\ref{fig:maps} and their specific properties are reported in Table~\ref{tab:photometry}.
The first section of the table presents the observations used for the global SED fitting, alongside the maps employed for the resolved SED fitting, with their resolutions and pixel sizes specified. The second section outlines the properties of ancillary maps used in various parts of this study.
The infrared/submm maps were retrieved from the DustPedia database\footnote{\label{dustpedia}\href{http://dustpedia.astro.noa.gr/}{http://dustpedia.astro.noa.gr/}} and the radio maps from the NASA/IPAC Extragalactic Database (NED)\footnote{\label{ned}\href{https://ned.ipac.caltech.edu/}{https://ned.ipac.caltech.edu/}}. In order to study the local physics of the galaxy at $\sim$~kpc spatial scales, and at the same time assuring sufficient wavelength coverage of the SED, we selected maps with resolution higher than 25$^{\prime\prime}$ corresponding to the SPIRE~-~350~$\mu$m beam.

Besides the maps and the photometric data that were used to constrain the SED modeling, we made use of several additional maps. In order to calculate the molecular gas content of the galaxy (see Sect.~\ref{subsec:ISM_morph}), but also to assess and correct for potential line contamination in the mm continuum bands (see Sect.~\ref{sec:process}), we made use of archival CO~(1-0) and CO~(2-1) maps retrieved from the BIMA SONG survey \citep{2003ApJS..145..259H} and HERACLES \citep{2009AJ....137.4670L}, respectively  (Table~\ref{tab:photometry}).
This study also incorporates supplementary \textit{Chandra} X-ray and \textit{Galaxy Evolution Explorer} (GALEX) far-ultraviolet (FUV) maps. The (Chandra-ACIS) X-ray map was retrieved from the \textit{Chandra} data archive\footnote{\label{chandra}\href{https://cxc.harvard.edu/cda/}{https://cxc.harvard.edu/cda/}} (PI: L. Jenkins). The GALEX FUV map was retrieved from the DustPedia database.

\section{Multi-wavelength maps processing} \label{sec:process}

The spatially resolved maps were processed and homogenized in a common way. The Homogenization of IMEGIN Photometry \citep[\texttt{HIP};][main developers: L.~Pantoni and J.~Tedros]{pantoniprep} pipeline, developed within the IMEGIN framework, accepts spatially resolved maps as inputs and homogenizes them in terms of resolution, grid, and units, performs the appropriate sky interpolation and subtraction, and accounts for the correct propagation of errors for each map.

The first step in the pipeline involves homogenization of the spectral flux density units of the various maps into Jy/px. Next, the CO~(2-1) line contamination is treated and subtracted from the NIKA2~-~1.15~mm map \citep[see NIKA2 transmission curves in][]{2020A&A...637A..71P}. The method that is followed in the pipeline is described in detail in \cite{2012MNRAS.426...23D}.
A spatially resolved CO~(2–1) map was used to compute the local contamination level across the galaxy. The resulting correction is not uniform: while the total contribution of the CO~(2–1) line to the global 1.15 mm emission is approximately 8\%, the contamination reaches $15-20$\% in the central nuclear region and decreases to about 3\% in the outer parts of the disk (see Fig.~\ref{fig:co21}).
We also tested for possible CO~(1–0) contamination in the 2~mm band and found that its contribution remains below 1\% even in the brightest CO regions. This level is negligible compared to our uncertainties, and no correction was applied.
Another step includes the removal of the brightest point sources in the field of the galaxy which are not associated with the galaxy itself. This is a crucial step, especially for NIR and MIR wavelengths, where foreground stars from the Galaxy may occupy a large portion of the observed frame.
The foreground sources were identified using the GAIA Data Release 3 catalog \citep{2022yCat.1355....0G} and masked out using circular apertures with a radius varying accordingly with the FWHM of the individual map. In order to protect the galaxy from over-subtracting sources in the area covered by the galaxy itself, a mask covering the main body of the galaxy was applied to the various maps. In the case of NGC~3627, this elliptical mask with a size of 200$^{\prime\prime}~\times~340^{\prime\prime}$, was centered at RA$_\mathrm{J2000}= 11^\mathrm{h}20^\mathrm{m}15^\mathrm{s}$ and DEC$_\mathrm{J2000} = +12^{\circ}59^{\prime}30^{\prime\prime}$, and rotated at an angle equal to the positional angle of the galaxy (173$^{\circ}$).

The residual sky emission was also subtracted from the auxiliary images. 
The sky emission is composed of several components, including the large-scale foreground emission, such as Galactic Cirrus in the MIR to submm wavelengths, residual fluctuations from the Cosmic Infrared Background (CIB) and the Cosmic Microwave Background (CMB) as well as any instrumental gradients that may be present. To isolate the relevant sky signals, the area associated with the galaxy was masked out (using the ellipse defined above). 
For the NIKA2 maps, only a small residual offset (less than 1\% of the total integrated flux) was fitted and subtracted, as the background fluctuations are removed during the data reduction.
The resulting map was then divided into a grid of cells, where the cell size could be adjusted based on the area covered by each map and the pixel size. The pipeline employs a sigma clipping algorithm to identify and manage outliers, specifically excluding the 5$\sigma$ detections. This step helps ensure that the presence of background/foreground sources that are detected in the map does not skew the estimated sky emission. Eventually, the sky emission computed in each cell is interpolated using a cubic spline, which smooths the background grid and enhances the accuracy of the background emission representation.

Finally, for the current analysis, we used {\tt HIP} to degrade the resolution of the maps to the SPIRE~-~350~$\mu$m beam (25$^{\prime\prime}$; 1.4~kpc at a distance of 11.32~Mpc). Given the point spread function (PSF) of each observation (see Table~\ref{tab:photometry}) each map was convolved with a Gaussian kernel in order to adjust the original PSF to the target PSF of 25$^{\prime\prime}$.
Following this step the maps were re-gridded into a common grid with a pixel size of 8$^{\prime\prime}$, which corresponds to a linear size of 0.44~kpc, and re-projected to the same orientation. A 3$\sigma$ threshold was also applied to the final maps.

To propagate uncertainties through each processing step, the pipeline uniquely employs a Monte Carlo (MC) method, in which all statistical uncertainties are sampled repeatedly to generate perturbed versions of the original map.
Any provided error map of the continuum emission or of the line emission, which is subsequently removed during line-contamination correction, is also incorporated into the propagated uncertainties.
For each MC iteration, a random normal distribution over the uncertainty range (centered at zero with a standard deviation equal to the estimated uncertainty at each pixel) is added to the original map, producing a perturbed map.
Each perturbed map is then processed in the same way as the original map.
Here, an iteration refers to an independent instance of the map with random perturbations.
After a specified number of iterations, a set of reprocessed maps is obtained. For the maps of NGC~3627, six MC iterations were used, as additional iterations were found to have no significant impact on the final result (less than 1\%).  
The final uncertainty at each pixel is computed as the standard deviation across the set of perturbed maps.
This approach allows for a more comprehensive treatment of uncertainties, ensuring a robust analysis.

Finally, the pipeline was used to measure the total emission of the galaxy within the abovementioned elliptical aperture positioned at the galaxy's center, so that it encompasses all the flux of the galaxy in every band.
The integration was carried out using the background-corrected map.
This ellipse was chosen so that it encompasses all the flux of the galaxy in every band. 
The global spectral luminosity $\nu L_{\nu}$ of the galaxy at 1.15~mm and 2~mm was found to be $(7.36~\pm~0.80)~\times~10^{6}$~L$_{\odot}$ and $(7.56~\pm~1.44)~\times~10^{5}$~L$_{\odot}$, respectively. Table~\ref{tab:photometry} presents the photometry values of the galaxy, measured in this study, along with additional literature values, which are included in the SED fitting.

\begin{figure}
    \centering
    \captionsetup{labelfont=bf}
    \includegraphics[width=0.5\textwidth]{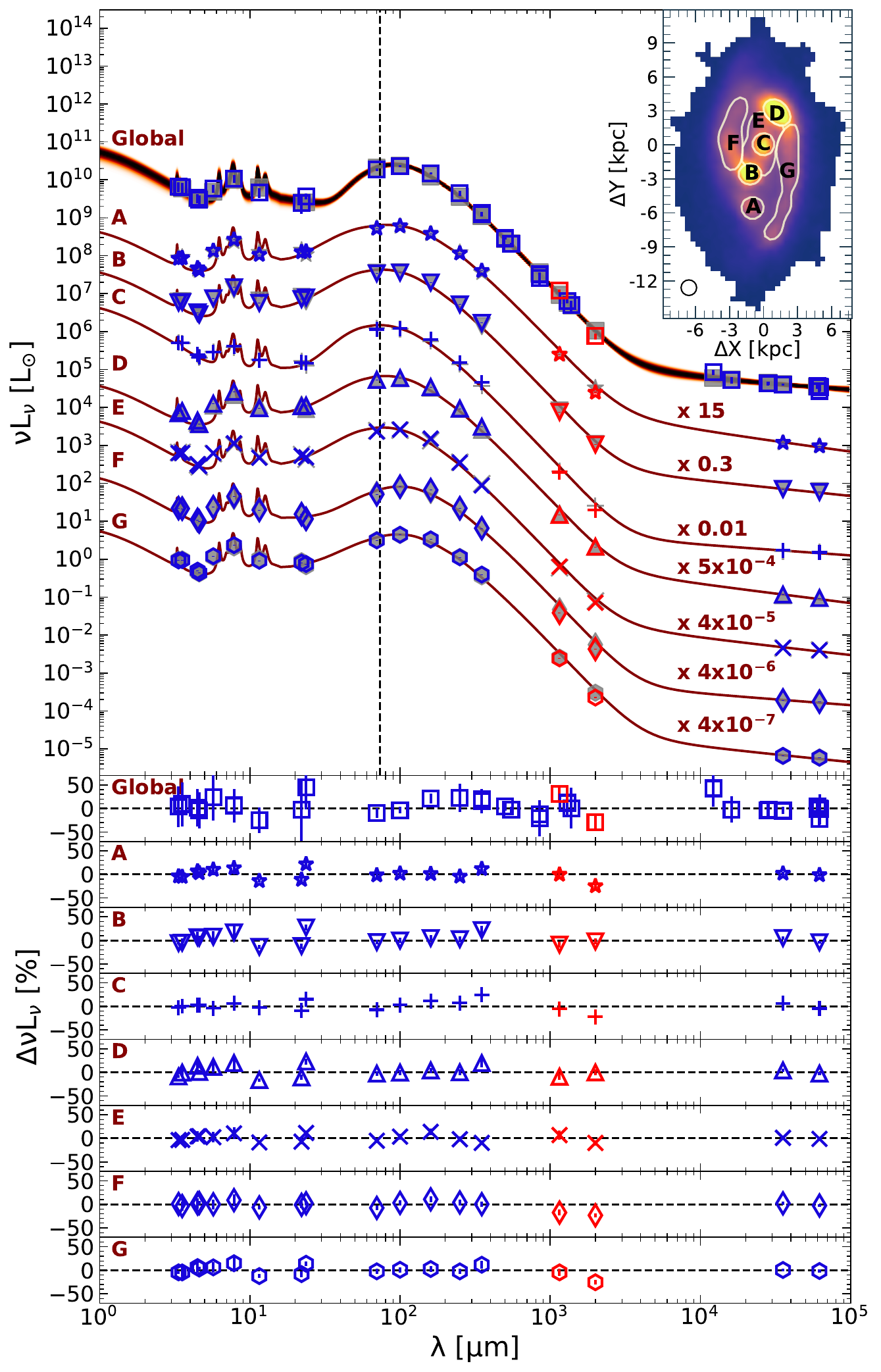} 
    \caption{ 
    The global SED of NGC~3627 followed by the SEDs of typical pixels within selected regions which correspond to an isolated \ion{H}{II} region, the southern bar-end, the nuclear region, the northern bar-end, the galactic bar, and the western and the eastern spiral arms of the galaxy (regions A to G, respectively). The SEDs are fitted with the \texttt{HerBIE} code (see Sect.~\ref{subsec:herbie} for a detailed description). 
    Observed spectral luminosities are represented by different blue symbols in each SED (with the exception of the NIKA2 measurements highlighted in red), while the corresponding, fitted models (and their uncertainties) are shown with continuous brown curves. 
    The gray filled symbols in each SED denote the model values at the observational bands. 
    The spectral luminosities for the global SED correspond to the values in the y-axis, while the rest of the SEDs are scaled by the factors indicated next to the model curves. 
    The vertical dashed line at 74.5~$\mu$m, marking the wavelength that the IR SED peaks for the galactic center (C), provides a reference of how the peaks of the other SEDs are positioned relative to each other. The respective residuals between observations and model are provided in the panels below.
    }
    \label{fig:seds}
\end{figure}

\section{SED fitting} \label{sec:fitting}

In order to constrain the properties of dust and radio emission in NGC~3627, we performed both global SED fitting, using integrated fluxes across all bands, as well as spatially resolved SED fitting at $\sim$~kpc scales, using a wealth of available multi-wavelength maps. This approach provides a comprehensive view of the galaxy’s emission properties from large to small scales. The analysis was conducted using the Hierarchical Bayesian Inference for dust Emission (\texttt{HerBIE}) SED fitting code \citep{2018MNRAS.476.1445G, 2021A&A...649A..18G}.

\subsection{SED fitting method} \label{subsec:herbie}

The \texttt{HerBIE} SED fitting code uniquely combines the treatment of several methodological aspects and a variety of sophisticated dust physical models. 
Spatially resolved SED analysis, using the  \texttt{HerBIE} code, has been applied by \cite{2023A&A...679A...7K} to the edge-on galaxy NGC~891.

\begin{table}
\begin{center}
\captionsetup{labelfont=bf} 
\caption{Global parameters derived from the SED fitting of NGC~3627 using the \texttt{HerBIE} code. }
\label{tab:global_sed}
\begin{tabular}{ll}
\hline \hline
 \textbf{Parameters}   & \textbf{Global values }  \\ \hline
    $M_\mathrm{dust}$~[M$_{\odot}$]        &  (2.05~$\pm$~0.09)~$\times$~10$^7$  \\
    $M_\mathrm{small~grains}$~[M$_{\odot}$]   &  (2.72~$\pm$~0.43)~$\times$~10$^6$
 \\
    $M_\mathrm{large~grains}$~[M$_{\odot}$]   &   (1.78~$\pm$~0.09)~$\times$~10$^7$
\\
    $M_\mathrm{gas}$/M$_\mathrm{dust}$   &  ~370~$\pm$~25  \\
    $T_\mathrm{dust}$~[K]      &  ~26.9~$\pm$~0.3
 \\
    $L_\mathrm{star}$~[L$_{\odot}$]       & (8.01~$\pm$~2.08)~$\times$~10$^{10}$  \\
    $L_\mathrm{dust}$~[L$_{\odot}$]       &  (3.95~$\pm$~0.19)~$\times$~10$^{10}$ \\
    $\langle U\rangle$~[$2.2\times10^5$~W~m$^{-2}$]    &   ~9.33~$\pm$~0.57 \\
    $q_\mathrm{AF}$          &   ~0.13~$\pm$~0.02
 \\
    $\beta$    &   ~1.76~$\pm$~0.43
   \\ 
    $\alpha_\mathrm{s}$      &  ~0.77~$\pm$~0.12
    \\ 
\hline
\end{tabular}
\end{center}

\begin{minipage}{\linewidth}
\small
\textbf{Notes.} The parameter $q_\mathrm{AF}$ represents the mass fraction of a-C(:H) particles smaller than 15~$\AA$ an analogous to $q_\mathrm{PAH}$ in the \cite{2007ApJ...657..810D} model.
The interstellar radiation field, described in \cite{1983A&A...128..212M}, is characterized by the average intensity $\langle U\rangle$, normalized to the solar neighborhood. Additionally, the Rayleigh-Jeans spectral index $\beta$ and the synchrotron spectral index $\alpha_\mathrm{s}$ are also included.
\end{minipage}
\end{table}

\texttt{HerBIE} employs the hierarchical Bayesian framework in order to determine the physical properties of a source that best explain its observed emission across a broad range of wavelengths. The physics of the dust grains are constrained by The Heterogeneous dust Evolution Model for Interstellar Solids \citep[\texttt{THEMIS};][]{2013A&A...558A..62J, 2017A&A...602A..46J}, a dust evolution model, based, the most, on laboratory-measured or laboratory-derived interstellar dust material properties. The model encompasses a diverse range of materials, including carbonaceous components, such as hydrogen-poor amorphous carbon (a-C) and hydrogen-rich amorphous carbon [a-C(:H)], each distinct in terms of structure, composition, and optical properties. In addition to carbon, \texttt{THEMIS} also incorporates silicate grains, which are abundant in interstellar environments and contribute significantly to the dust composition. The model considers various dust components, including amorphous silicates, and accounts for the heterogeneous nature of dust and its evolution through processes such as growth, fragmentation, and destruction. By applying a distribution of starlight intensities \citep{2001ApJ...549..215D} to account for the stochastic heating of the interstellar dust material \citep{1989ApJ...345..230G}, our approach allows for a more realistic representation of how dust grains, of various sizes and composition, interact with different radiation fields within galaxies. In addition to the IR/submm thermal dust emission, \texttt{HerBIE} also treats the emission at radio wavelengths, accounting for both free-free emission and synchrotron emission. 
For our analysis we used the {\fontfamily{qcr}\selectfont{powerU}}, {\fontfamily{qcr}\selectfont{starBB}} and {\fontfamily{qcr}\selectfont{radio}} SED modules [see Sect.~2.2 in \cite{2018MNRAS.476.1445G} and Sect.~3 in \cite{2021A&A...649A..18G}] resulting in a total of ten free parameters per pixel. 

Although NGC~3627 is classified as hosting a Seyfert-2 AGN, its contribution to the radio continuum appears negligible. VLBA observations at 6~cm \citep{2004A&A...418..429F} and estimates at 9~mm \citep{2015ApJ...813..118M} suggest the AGN accounts for less than 3\% of the nuclear emission. We therefore do not include an AGN component in the SED modeling and assume the nuclear emission is primarily powered by star formation.

Additionally, \texttt{HerBIE} allows the extent of the prior distribution to the external parameters, a technique shown to efficiently recover the true correlations between the fitted and external parameters \citep[see Sect.~5.3 in][]{2018MNRAS.476.1445G}. Comparing the different priors, this technique improves the fitting performance of a large sample of sources in low signal-to-noise environments \citep[see Sect.~V.3.3.2 in][]{2022HabT.........1G}. In this study, we used the spatial distribution of atomic and molecular hydrogen as prior external knowledge in the fitting process. 
Furthermore, the fitting code accounts for color corrections and the inclusion of the calibration uncertainties in each band (see Table~\ref{tab:photometry}).

\subsection{SED fitting of NGC~3627} \label{subsec:seds}

\begin{figure*}
    \centering
    \captionsetup{labelfont=bf}
    \includegraphics[width=\textwidth]{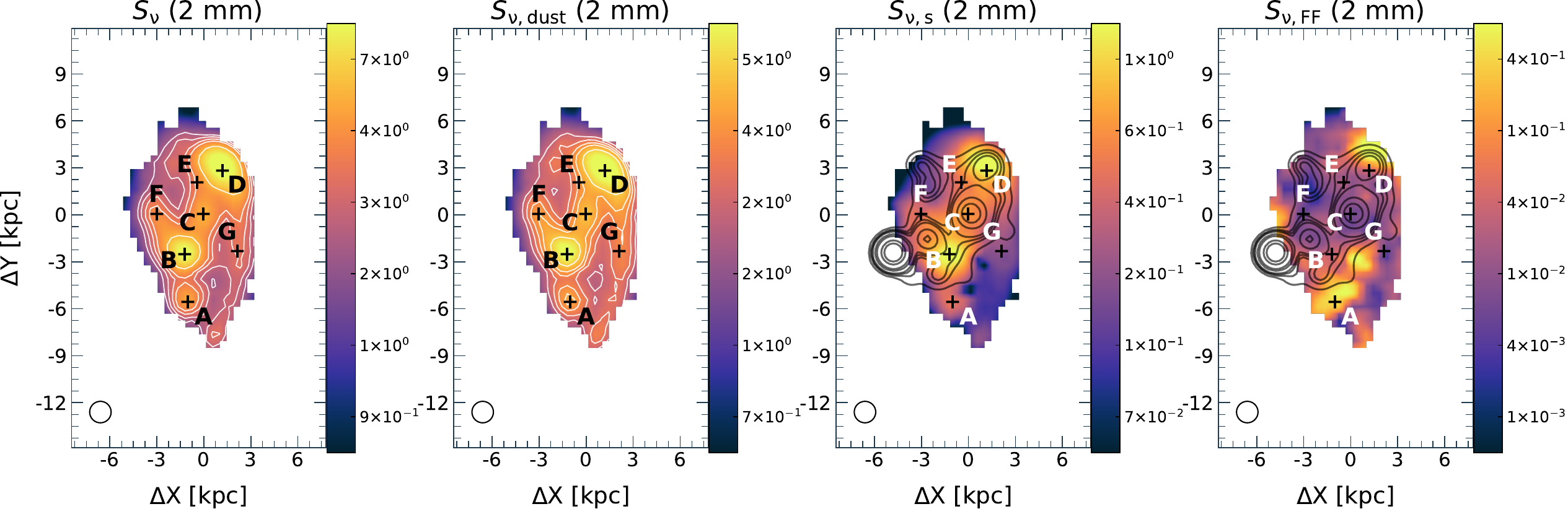}
    \caption{Total modeled emission at 2 mm (left panel) and its decomposition into dust, synchrotron, and free-free emission, shown in the second, third, and last panels, respectively. 
    The color bars are in mJy per 25$^{\prime\prime}$ beam, and the beam size is indicated in the bottom-left corner in each frame. 
    The crosses indicate the regions A to G (see Fig.~\ref{fig:seds}). Contours of X-ray emission (observed by \textit{Chandra}) have been superimposed on the synchrotron and free-free emission maps, while the white contours in the first two panels are employed to aid in visualizing the galaxy's morphology. As proposed by \citet{2012A&A...544A.113W}, the asymmetric X-ray extension in the eastern part of the galactic disc may be linked to a recent interaction with a dwarf galaxy approaching from this direction.
    }
    \label{fig:2mm}
\end{figure*}

In order to probe the global properties of the galaxy, we modeled the integrated spectral luminosity $\nu L_{\nu}$ from 3.4~$\mu$m to 6.2~cm (in 33 bands) of the galaxy, reported in Table~\ref{tab:photometry} and measured as described in Sect.~\ref{sec:process}. 
Taking advantage of the high-resolution capabilities of instruments observing at the IR to radio wavelengths (see Sect.~\ref{sec:data}), 
we also fitted the resolved spectral luminosity of the galaxy pixel-by-pixel, which was possible for 18 of those 33 bands. 
After processing the maps as described in Sect.~\ref{sec:process} we fitted the SEDs in each pixel (covering an area of a $0.44~\times~0.44$~kpc$^2$). The usable data for the pixel-by-pixel SED fitting were the photometric pixel values of the maps that exceeded the 3$\sigma$ threshold, with an additional constraint requiring at least four measurements within the FIR/submm range. The number of the available photometric data led to a global SED fitting with 23 degrees of freedom and to a spatially resolved SED fitting with 4–8 degrees of freedom. As external priors of the spatial fit, we also included the maps of \ion{H}{I} and CO~(1-0) line emission (see Fig.~\ref{fig:maps} for their distribution, and Table~\ref{tab:photometry} for their properties) as tracers of the atomic and molecular gas distributions, respectively.

In Fig.~\ref{fig:seds} we have plotted the global SED of the galaxy, followed by the SEDs in pixels of characteristic regions of the galaxy noted with letters from A to G. The SPIRE~-~350~$\mu$m map has been used as a reference to distinguish the galactic regions. Region A is a large, isolated \ion{H}{II} region located about 5.7~kpc from the center (see the inset map in Fig.~\ref{fig:seds}). Regions B and D are the bright bar-ends (the regions where the bar stops and are connected to the spiral arms) on the southern and northern parts of the galaxy, respectively. Region C is the nucleus of the galaxy, with region E encompassing the bar itself. Regions F and G are representative of the spiral arms, including a large part of the eastern and western spiral arms, respectively. The seven regions, A to G, cover areas of 2, 7, 3.1, 2.7, 4.6, 9.3, 11.9, and 13.5~kpc$^2$, respectively. Regarding the quality of the fit of the model to the data, the global SED is characterized by a reduced-$\chi^2$ value of 1.43 while the reduced-$\chi^2$ values of the spatially resolved SED fitting are below 20 for 84\% of the 1465 fitted pixels with a mode value of $\chi^2=3.9$ (see Fig.~\ref{fig:chi2}).

In Fig.~\ref{fig:seds}, the vertical dashed line marks the wavelength where the peak dust emission occurs in region C. It is clear that this peak, found at 74.5~$\mu$m, is positioned more at shorter wavelength compared to other SEDs. This indicates higher dust temperatures, as is anticipated for the nuclear region C ($\sim$~31~K), compared to the rest of the regions ($28-30$~K in the bar-ends B and D), and especially for regions F and G ($24-26$~K) that show a peak emission occurring at larger wavelengths. We note that the 1$\sigma$ uncertainties on the temperature estimates are less than 0.1~K per pixel. 

\subsubsection{Global SED fitting of NGC~3627}\label{ssubsec:global}

Table~\ref{tab:global_sed} includes the values of the physical properties derived from the global SED analysis of the galaxy. 
To estimate the gas mass, we utilized the available VLA/\ion{H}{I} and BIMA/CO~(1-0) maps to determine the masses of atomic and molecular gases, respectively (refer to Table~\ref{tab:photometry}). For these computations, we employed the recipes outlined in \cite{2008AJ....136.2563W} and \cite{2011A&A...527A..92C} for atomic and molecular gas mass calculations, respectively, which are detailed in Section~\ref{subsec:ISM_morph}.
These calculations indicate an atomic gas mass of $M_\mathrm{HI}=1.10\times10^9$~M$_{\odot}$, and a molecular gas mass of $M_\mathrm{H_2}=4.48\times10^9$~M$_{\odot}$ resulting in a total gas mass (allowing for a helium contribution of 1.36) of $M_\mathrm{gas}=7.59\times10^{9}$~M$_{\odot}$. A more detailed description of these conversions is detailed in Sect.~\ref{sec:ISM}. Apart from the global SED fitting results, Table~\ref{tab:spatial_sed} presents the median values of the estimated physical properties obtained from the spatially resolved SED fitting for the regions A to G.

Our analysis indicated a total dust mass of 2.05~$\times$~10$^7$~M$_{\odot}$ for NGC~3627, and a corresponding average dust temperature of 26.9~K. 
Previous studies have performed global SED fitting using different models, generally estimating a higher dust mass than our results. \cite{2012MNRAS.425..763G} modeled the MIR to submm emission with both a modified blackbody (MBB) and the \cite{2007ApJ...657..810D} model, obtaining dust masses of 1.3~$\times$~10$^8$~M$_{\odot}$ and 6.4~$\times$~10$^7$~M$_{\odot}$, respectively. 
More recently, \cite{2023AJ....165..260D}, using the \cite{2014ApJ...780..172D} dust model prescription, reported a dust mass of 8~$\times$~10$^7$~M$_{\odot}$.
These differences largely reflect the dependence of dust mass estimates on the adopted dust model \citep[see, e.g.,][]{2021ApJ...912..103C}.
Variations in grain optical properties, emissivity assumptions, and the treatment of the Rayleigh–Jeans tail can shift the absolute dust mass by up to a factor of $\sim$3 across commonly used dust models.
Similarly, \cite{2019A&A...624A..80N} and \cite{2020MNRAS.496.3668D} fitted the global SED using the \texttt{THEMIS} dust model, reporting dust masses of 3.8~$\times$~10$^7$~M$_{\odot}$ and 2.9~$\times$~10$^7$~M$_{\odot}$, respectively. All values have been rescaled to match the galaxy distance adopted in this study. 
The difference in dust mass estimation could stem from the inclusion of NIKA2 photometric data in our analysis, which provides additional constraints in the mm regime, as
well as from the use of the latest photometric measurements conducted with \texttt{HIP}, and/or the use of \texttt{HerBIE}. 
The derived dust mass leads to a gas-to-dust mass ratio of 370. This value is consistent with the expected trend for subsolar metallicity environments \citep[see, e.g.,][]{2014A&A...563A..31R}, given a mean gas metallicity of 0.85~Z$_{\odot}$ for NGC~3627 \citep{2022MNRAS.509.1303W}.

In the global SED fitting of the galaxy, we derived a synchrotron spectral index of 0.77~$\pm$~0.12, which is consistent, within uncertainties, with the value of 0.89$^{+0.22}_{-0.15}$ found by \cite{2017ApJ...836..185T} for NGC~3627. This spectral index reflects the typical power-law behavior of synchrotron radiation, produced by relativistic electrons spiraling in magnetic fields. 
Its determination is primarily constrained by data at radio wavelengths ($\lambda~\geq$~1~cm), where synchrotron emission dominates the SED, although mm measurements may also contribute when radio sampling is limited \citep[see, e.g.,][]{2023A&A...679A...7K, 2025A&A...693A..88E}.
While such a value is indicative of ongoing star formation—where supernova explosions from massive stars accelerate cosmic rays that contribute to synchrotron emission \citep[e.g.,][]{2015MNRAS.449.3879B, 2017A&A...602A...6S, 2023MNRAS.525.3413C}—the presence of an AGN in the galaxy’s center may contribute modestly. 
However, given that most Seyfert 2 galaxies are radio-quiet and lack prominent jets \citep[e.g.,][]{2020MNRAS.499..334S}, the AGN is unlikely to dominate the synchrotron spectrum here.

In the submm/mm regime, the discrepancy between the global SED model and the observations does not support the presence of excess emission, as reported by \cite{2014MNRAS.439.2542G}. However, at 1.2~cm, the detected emission exceeds the model's prediction, similar to the findings of \cite{2022A&A...658L...8B}, which may be attributed to AME. 
We note that the \texttt{HerBIE} model setup used in this study does not include an explicit AME component. We have tested alternative fits including an AME component and found that the excess remains within the model uncertainties. Nonetheless, the current data do not allow us to robustly assess the presence of AME. Follow-up observations in the $5~\mathrm{mm}-3~\mathrm{cm}$ range would be essential to confirm the nature of this excess and to robustly disentangle potential AME from other emission mechanisms.

\begin{figure}
    \centering
    \captionsetup{labelfont=bf}    \includegraphics[width=.41\textwidth]{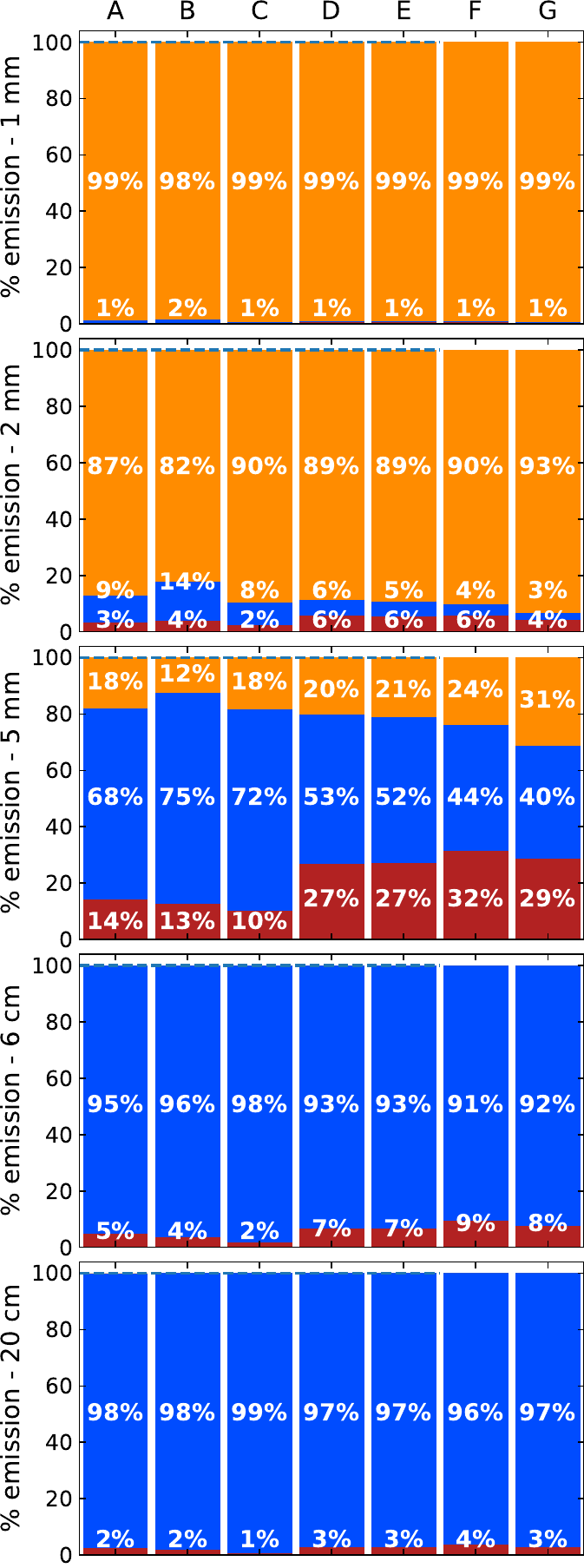} 
    \caption{Emission components contributing to the total flux at 1, 2 and 5~mm, 6 and 20~cm (top to bottom panels) in the regions A, B, C, D, E, F and G (left to right).
    The emission percentages for the dust, the free-free, and the synchrotron emissions are shown in orange, red, and blue bars, respectively. 
    The actual percentage values for each emission mechanism, in each region, are indicated with numbers in the plots. 
    }\label{fig:barplots_regions}
\end{figure}

\section{Emission components at millimeter wavelengths} \label{sec:decomposition}

The emission powered up by the complex mm spectral range consists of several mechanisms, which must be carefully disentangled. It is thus interesting to determine how the total emission is formed in various environments throughout the galaxy. In this study, we focus on the model-derived contributions to the emission sources at mm/cm wavelengths. As the derived model represents the best-fit interpretation of the available observations, it offers a systematic way to extract and analyze the individual components.

\subsection{Decomposition of the 2~mm model emission}
In the leftmost panel of Fig.~\ref{fig:2mm} the modeled emission map at 2~mm is presented and decomposed, using the \texttt{HerBIE} model, into its dust emission, synchrotron emission, and free-free emission components. In all panels, the positions of the regions of interest, A to G, are indicated with crosses, while the open circle, in the bottom-left corner, indicates the 25$^{\prime\prime}$ resolution of the final modeled maps.
Contours of X-ray emission traced by \textit{Chandra} observations have also been over-plotted on the synchrotron and free-free emission maps. The modeled maps shown here are generated using a more strict masking criterion, namely, present regions with spectral luminosities at NIKA2 and VLA wavelengths exceeding the 3$\sigma$ threshold. This selection ensures that the modeled emission is constrained by adequate observational data (given that only two radio wavelengths are available at 3.5 and 6.2~cm), reducing uncertainties in regions with weak or poorly sampled emission.
 
Firstly, it is worth noticing here that the general morphology of the model (leftmost panel) is similar to the actual 2~mm observation, as expected, with the very prominent bar-ends (B, and D) connected by the bar (E) with a fainter nucleus at the center of the galaxy (C) and a relatively bright, isolated, \ion{H}{II} region (A). The spiral arms F, and G, are also obvious and originating from their starting point, the bar-ends B, and D, respectively. The dust emission (second panel) being the dominant component of the galaxy at 2~mm  follows the morphology of the global emission as described above with only minor differences, not easily detectable by eye, in the bright parts of the galaxy.

Synchrotron emission (third panel) becomes the second most prominent contributor to the total emission at 2~mm in this galaxy. Notably, there is enhanced emission at the two bar-ends (regions B and D), and the bar itself (region E) along with the nucleus (region C) appears relatively bright.
The synchrotron emission is relatively weak in the spiral arms (regions F and G). 
The isolated \ion{H}{II} region (A) exhibits noticeable synchrotron emission, though at lower levels compared to the two bar-ends.

The free-free emission map (rightmost panel), on the other hand, shows a more complex and unexpected behavior with maximum emission not directly linked to the bright bar-ends (regions B and D) but rather distributed in a more homogeneous way, spanning a small dynamic range apart from a few exceptions. One such exception is region A, the isolated \ion{H}{II} region in the southern part of the galaxy, which shows an obvious peak in free-free emission, as expected.
Hot, young stars are the sources of the ionized gas and thus the observed free-free emission. Apart from this region, there are two more regions, one between regions B and G and one in the bar-end D which also show enhanced free-free emission. If our modeled decomposition is accurate, these regions trace ionized gas. Due to the low presence of dust in these regions, gas shielding is relatively low, allowing UV radiation to ionize gas emitting free-free radiation.

To further investigate the derived free-free emission, we compared it with the X-ray radiation emitted by the galaxy. As illustrated by the contours of the two right panels, X-ray emission peaks in the nucleus and the eastern part of the galactic disk, with additional secondary peaks near the northern bar-end (region D) and just north of the southern bar-end (region B). \cite{2012A&A...544A.113W} linked the extension of X-ray emission to the east with a recent collision with a dwarf galaxy approaching from that direction. We propose that strong shocks driven by the two eastern ultra-luminous X-ray sources (ULXs) influence the ionized gas dynamics, leading to a localized free-free deficit at their peaks. At the same time, these shocks likely compress and heat the surrounding gas, enhancing free-free emission in neighboring regions. 
This scenario is in agreement with theoretical and observational studies of ULXs \citep[e.g.,][]{2016MNRAS.455L..91A}, where the presence of high X-ray luminosity paired with faint optical counterparts is interpreted as evidence for super-Eddington accretion onto stellar-mass black holes. In such cases, powerful disc winds and collimated outflows are expected to drive shocks and reshape the surrounding interstellar medium.
We caution, however, that due to the small number of wavelengths available to constrain the radio emission, and due to the fact that the free-free emission is low ($\sim2-5$\%) and thus, within the uncertainties of the model, the free-free emission may not be reliably calculated.
Additional observations at intermediate wavelengths (e.g., $2~\mathrm{mm}-3~\mathrm{cm}$) with comparable spatial resolution would be valuable for better constraining the free-free component and assessing its physical validity.

\subsection{Exploring the mm-to-cm emission components}

To investigate the scarcely explored mm-to-cm spectral range, Figure~\ref{fig:barplots_regions} presents the contributions of different emission mechanisms to the total modeled emission of NGC~3627 at 1~mm, 2~mm, 5~mm, 6~cm, and 20~cm across the selected regions (A to G). From these bar plots, it is evident that at 1~mm the emission is purely coming from dust. At 2~mm, as previously discussed, radio emission begins to emerge, with a minor free-free contribution and a more pronounced synchrotron component. Notably, in region B (the southern bar-end), synchrotron already accounts for 14\% of the total flux.
At 5~mm, thermal dust emission declines (though it still contributes $\sim20$\%), while synchrotron emission rises sharply, dominating $40-75$\% of the total flux depending on the region. At this stage, where radio components become increasingly significant, clear differences emerge between regions A to C—closer to the highest X-ray peaks—and the other regions (D to G). In particular, in the central and eastern regions (A to C), synchrotron contributes $\sim70$\% of the radio emission at 5~mm, with free-free emission at $\sim12$\%.
In contrast, at the galactic bar (E) and the northern bar-end (D), synchrotron contributes slightly less ($52-53$\%), while free-free emission rises to 27\%, and cold dust still accounts for 20\%. The spiral arms (F and G) exhibit the highest contribution from free-free and cold dust emission.

At longer wavelengths, such as 6~cm, the emission is purely radio, with synchrotron dominating ($>90$\%) across all regions. 
Free-free emission is negligible in the nucleus but contributes $\sim5$\% in regions A and B, $\sim7$\% in the galactic bar and the northern bar-end, and slightly higher ($8-9$\%) in the spiral arms. At 20~cm, synchrotron radiation prevails across all regions, with free-free emission contributing less than 5\%.

On a global scale, (87~$\pm$~5)\% of the total 2~mm emission in NGC~3627 originates from cold dust, while synchrotron and free-free emission contribute (6~$\pm$~2)\% and (7~$\pm$~5)\%, respectively. This is consistent with the findings of \cite{2023A&A...679A...7K} for the edge-on star-forming galaxy NGC~891, where an analysis using the same methodology showed that 91\% of the 2~mm emission comes from cold dust, with free-free and synchrotron emission contributing 5\% and 4\%, respectively. The slightly higher contribution of radio components in NGC~3627 may arise from several mechanisms, including synchrotron emission associated with its Seyfert-2 AGN, star formation-related processes, or shock-driven emission potentially linked to the nearby ULXs. 
Additionally, the integrated radio SED of NGC~3627 has been previously analyzed by \cite{2017ApJ...836..185T}, who found that free-free emission accounts for 16\% at 6~cm and 6\% at 20~cm. Our integrated SED fitting yields comparable fractions within their uncertainties, predicting that ($9~\pm~7$)\% of the total emission at 6~cm originates from free-free emission, decreasing to ($4~\pm~3$)\% at 20~cm.

\begin{figure}
    \centering
    \captionsetup{labelfont=bf}    \includegraphics[width=0.5\textwidth]{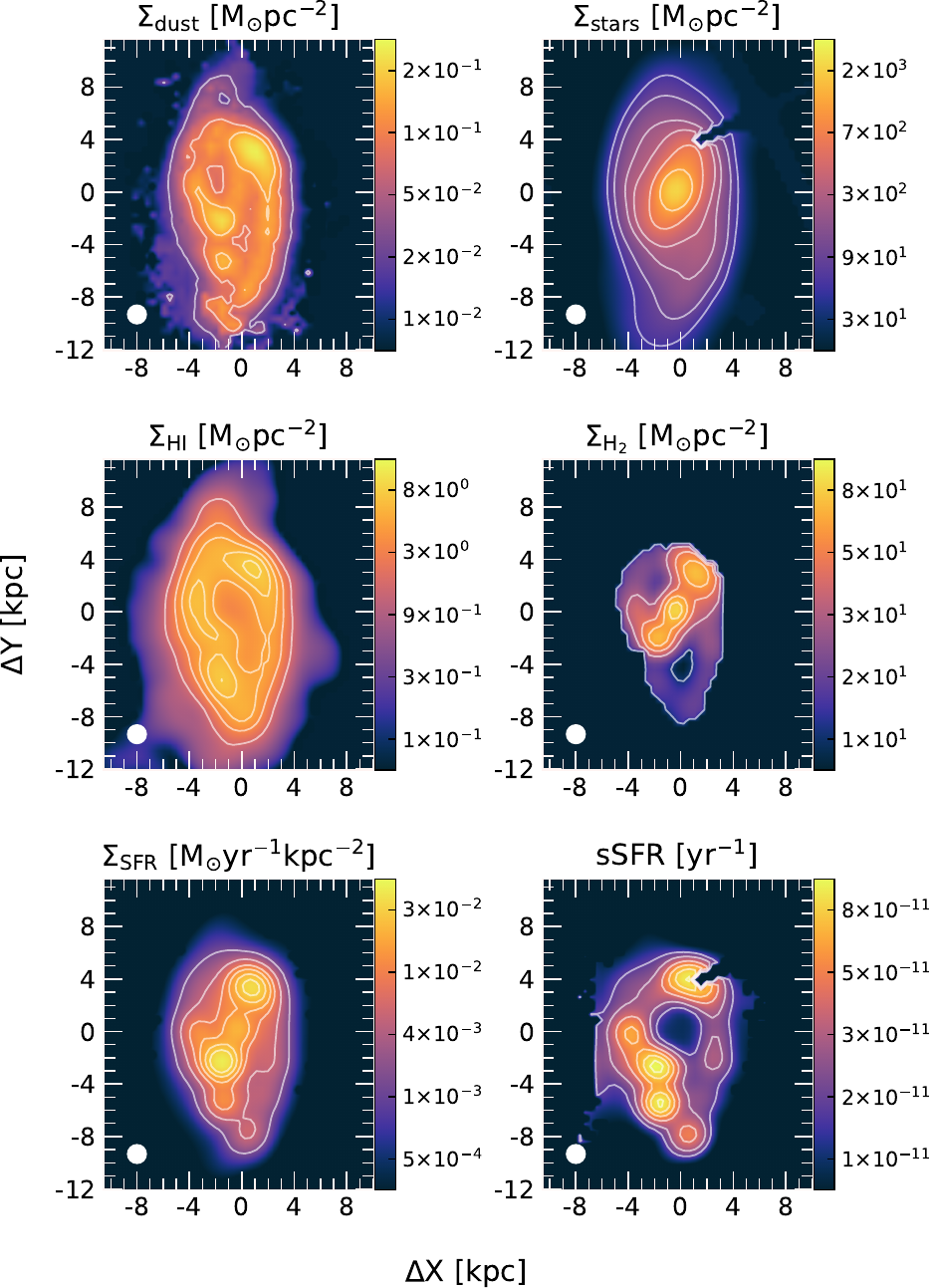} 
    \caption{
    The dust mass and stellar mass surface densities (top panels; left and right, respectively), the atomic mass and molecular mass surface densities (middle panels; left and right, respectively), and the  star formation rate (SFR) and specific star formation rate densities (bottom panels; left and right, respectively). All maps share the same grid and resolution of 25$^{\prime\prime}$ (indicated by the white circle in the lower left corner). The methods used in deriving these maps are detailed in Sect.~\ref{subsec:ISM_morph}. 
    }
    \label{fig:components}
\end{figure}

\begin{figure*}
    \centering
    \captionsetup{labelfont=bf}    \includegraphics[width=\textwidth]{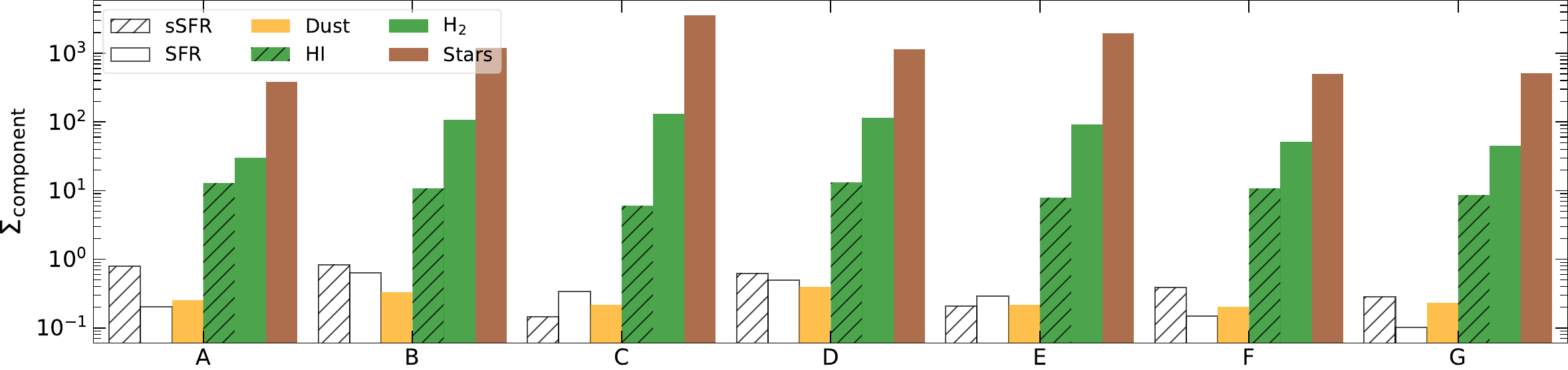} 
    \caption{
    The ISM and stellar mass content, as well as SFR and sSFR 
    indicators, included in each region of interest (A to G) in NGC~3627. The colored bars indicate the average surface densities (within each region) of dust mass (in orange), atomic gas mass (in hatched green), molecular gas mass (in solid green), and stellar mass (in brown) expressed in units of M$_{\odot}$~pc$^{-2}$.
    Additionally, the average star formation rate (SFR; in hatched white) and specific star formation rate (sSFR; in plain white) densities, are included in this plot. SFR, and sSFR are given in units of M$_{\odot}$~yr$^{-1}$~kpc$^{-2}$ and yr$^{-1}$, and scaled by factors of 10 and 10$^{10}$, respectively.
    }
    \label{fig:compont_barplt}
\end{figure*}

\section{The ISM in NGC~3627}\label{sec:ISM}

NGC~3627 is a very interesting galaxy to study for a number of reasons. It has a distinct barred structure with tightly wound asymmetric spiral arms, which are likely to be formed like this due to past gravitational interactions with its neighboring galaxies in the Leo Triplet. The galaxy contains regions of enhanced star formation, particularly in its bar-ends, but also in regions associated with the spiral arms. Finally, the galaxy is classified either as a Seyfert 2 AGN or as a LINER, indicating its nucleus exhibits mild activity, possibly driven by a supermassive black hole. These characteristics have a direct impact on the ISM and certainly warrant detailed investigation. Given the limited spectral diagnostics available, we adopt the working assumption that the nuclear emission is primarily powered by star formation throughout the rest of our analysis.

\subsection{The ISM morphology and abundances and its relation to SFR}\label{subsec:ISM_morph}

For a comprehensive understanding of the ISM and its constituents, and besides the dust mass map which has already been calculated through the SED modeling (see Fig.~\ref{fig:components}, top-left panel), we produce maps for the stellar mass, atomic hydrogen mass, molecular
hydrogen mass, as well as the SFR and the specific star formation rate (sSFR).
All these maps are expressed in column density or surface density units, which have been corrected for inclination. The observational tracers of these parameters (see the following equations) were processed using the \texttt{HIP} pipeline, as described in Sect.~\ref{sec:process}.

As shown in the top-left panel, the highest dust concentration is found in the northern bar-end (region D), which contains more than twice the dust mass of the southern bar-end (region B). The two bar-ends together account for 20\% of the total dust mass, while region A, a prominent isolated star-forming site, contains 6\% of the total dust mass. Finally, the nucleus and the galactic bar (regions C and E) contain 2\% and 8\% of the total dust mass, respectively. In total, 36\% of the dust mass of the galaxy is confined in the distinct and brighter regions of the galaxy, with the remaining 64\% distributed in the spiral-arm and inter-arm regions. 

The stellar mass (see Fig.~\ref{fig:components}, top-right panel) was calculated using the method proposed by \cite{2012AJ....143..139E} using the IRAC 3.6~$\mu$m and 4.5~$\mu$m maps, mainly tracing the old stellar populations. The calibration assumes a Salpeter initial mass function \citep[IMF;][]{1955ApJ...121..161S}. The equation is as follows:
\begin{equation}
    M_{\star}[\text{M}_{\odot}] = 10^{5.65}~(F_{3.6}[\text{Jy}])^{2.85}~(F_{4.5}[\text{Jy}])^{-1.85}~(\frac{D[\text{Mpc}]}{0.05})^2.
    \label{eq:starmass}
\end{equation}
The 
This expression is calibrated for the Large Magellanic Cloud (LMC) and is applicable to a variety of environments. 
However, as the authors point out, the greatest discrepancies between flux and stellar mass tend to appear in areas with younger stellar populations ($<300$~Myr) and where hot dust is present, indicated by 8~$\mu$m emission.
The derived map reveals a pronounced nucleus as the brightest part throughout the surface of the galaxy, and a stellar bar is evident, extending roughly 2.8~kpc on each side of the nucleus. 
The spiral arms' stellar structure is also discernible, but exhibits a considerably lower surface density compared to the nucleus and the bar. Moreover, an asymmetry between the spiral arms is apparent, with the southern arm being significantly more extended than the northern one.

\begin{figure*}
    \centering
    \captionsetup{labelfont=bf}    \includegraphics[width=\textwidth]{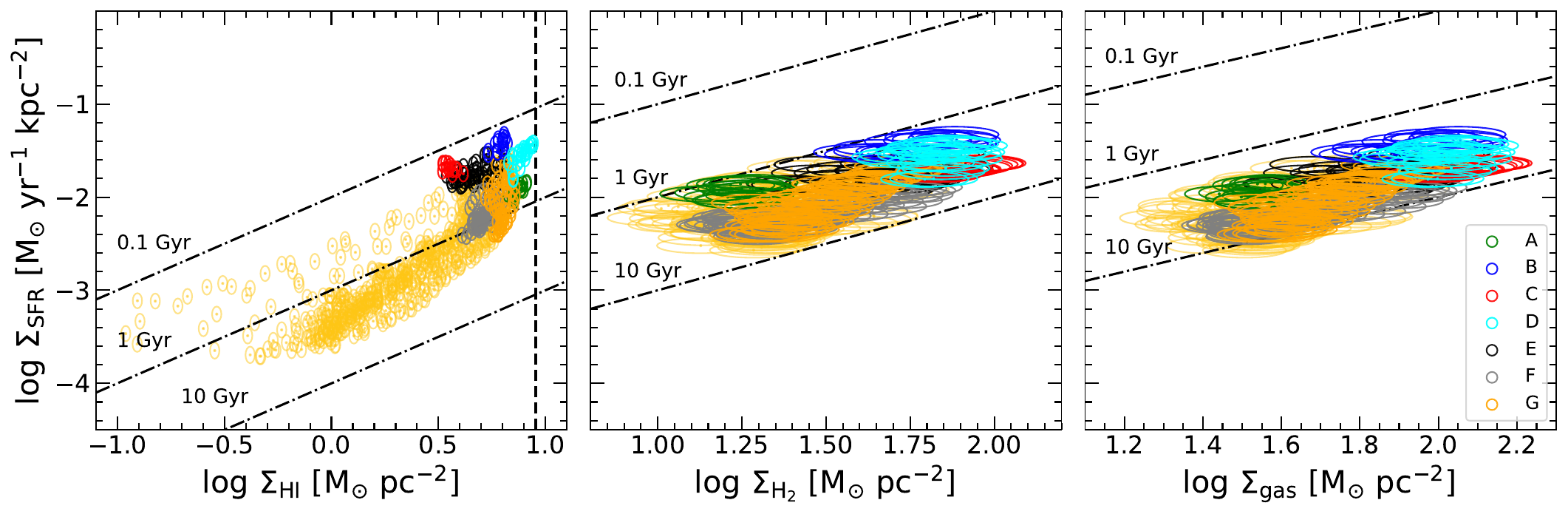} 
    \caption{The Kennicutt-Schmidt relation for NGC~3627, for the atomic, the molecular and the total gas mass (left to right panels, respectively). The maps presented in Fig.~\ref{fig:components} are used in this plot. Each data point represent one pixel in these maps covering an area of 8$^{\prime\prime}$~$\times$~8$^{\prime\prime}$ (equivalent to 0.44~$\times$~0.44~kpc$^{2}$). Different colors represent the regions from A to G, as indicated in the inset panel of the rightmost plot, with the light yellow data points representing the remaining mapped parts of the galaxy. The vertical dashed line denotes the saturation threshold of $\Sigma_{\mathrm{HI, saturation}}=9~\mathrm{M}_{\odot}$~pc$^{-2}$, as proposed by \citet{2008AJ....136.2846B}.}
    \label{fig:ken-schm}
\end{figure*}

The mass of atomic hydrogen was determined using the VLA/\ion{H}{I} map (see Table~\ref{tab:photometry}).
Employing the equation from \cite{2008AJ....136.2563W} on the \ion{H}{I} line emission $\sum_i S_i \Delta\upsilon$ (total emission summed over all channels), we computed the atomic hydrogen mass as follows:
\begin{equation}
    M_\mathrm{HI} \left[\text{M}_{\odot}\right] = 2.36 \times 10^5~(D[\text{Mpc}])^2 \times \sum_i S_i \Delta\upsilon \left[\frac{\text{Jy}} {\text{km} \ \text{s}^{-1}}\right].
    \label{eq:HI}
\end{equation}
The map of the atomic hydrogen is shown in Fig.~\ref{fig:components} (middle-left panel) revealing a ring-like structure tracing the spiral arms of the galaxy. 
The atomic hydrogen ring structure is particularly bright along the spiral arms, with extended but much fainter emission also associated with a diffuse component, especially in the outer regions of the galaxy. Notably, there is a clear deficit of atomic hydrogen in the central region and along the bar. 
However, two prominent bright regions stand out within the ring-like structure: one coincides with the northern bar-end (region D), while the other is linked to the star-forming region A in the southern part of the galaxy. Atomic hydrogen column density appears to be lower in region B, corresponding to the southern bar-end, compared to that toward region D.

The mass of molecular gas, as depicted in the middle-right panel of Fig.~\ref{fig:components}, has been determined utilizing the BIMA/CO~(1-0) map along with equation (1) from \cite{2011A&A...527A..92C}:
\begin{equation}
    M_{\mathrm{H}_2}[\text{M}_{\odot}]=8.653\times10^3~(D[\text{Mpc}])^2~S_\mathrm{CO(1-0)}[\text{Jy}~\text{km}~\text{s}^{-1}].
\end{equation}
This equation adopts a standard CO-to-H$_2$ conversion factor of $X_\mathrm{CO}=2.2\times10^{20}~\mathrm{cm}^{-2}~(\mathrm{K~km}~\mathrm{s}^{-1})^{-1}$, consistent with the calibration of \citet{1991IAUS..146..235S}. Since this factor depends on various physical conditions, such as metallicity, we apply a 30\% uncertainty on the derived values to account for associated systematics \citep[e.g.][]{2013ARA&A..51..207B}.

To construct the SFR map (Fig.~\ref{fig:components} bottom-left panel), we combine the dust emission at 24~$\mu$m, which is predominantly heated by young massive stars, with the dust-obscured FUV emission.
The calibration formulated by \cite{2011ApJ...741..124H} utilizes MIPS~-~24~$\mu$m and GALEX~-~FUV observations to calculate the SFR through the equation:
\begin{equation}
\text{SFR}[\text{M}_{\odot}\text{yr}^{-1}] = 3.36 \times 10^{-44}~(L_\mathrm{FUV} [\text{ergs}^{-1}] + 3.89~L_\mathrm{24\mu m} [\text{ergs}^{-1}]).
\end{equation} 
This calculation assumes a Kroupa IMF \citep{2003ApJ...598.1076K}. 
The specific SFR map (Fig.~\ref{fig:components} bottom-right panel) can then be directly computed by dividing SFR with the stellar mass:
\begin{equation}
    \mathrm{sSFR}~[\mathrm{yr}^{-1}] = \frac{\mathrm{SFR}~[\mathrm{M}_{\odot}~\mathrm{yr}^{-1}]}{M_{\star}~[\mathrm{M}_{\odot}]}.
\end{equation}
As the stellar mass was originally derived assuming a Salpeter IMF, we converted it to a Kroupa IMF by applying a scaling factor of 0.67, following \cite{2014ARA&A..52..415M}.
The SFR surface density map shows the current star formation activity of the galaxy, which is extremely high in the bar-ends (regions B and D). It is of interest to notice that, although between the two bar-ends it is the northern one that is brighter in all components of the ISM (dust, atomic and molecular hydrogen), the southern one (region B) shows higher $\Sigma_{\mathrm{SFR}} \sim  0.07$~M$_{\sun}~\text{yr}^{-1}~\text{kpc}^{-2}$ (compared to $\sim  0.05$~M$_{\sun}~\text{yr}^{-1}~\text{kpc}^{-2}$ for region D). The next more active region of the galaxy, in terms of star formation, is the nucleus ($\sim  0.03$~M$_{\sun}~\text{yr}^{-1}~\text{kpc}^{-2}$) followed by the isolated star-forming region A ($\sim 0.02$~M$_{\sun}~\text{yr}^{-1}~\text{kpc}^{-2}$). The spiral arms show a bit lower $\Sigma_{\mathrm{SFR}}$ of $\sim 0.01$~M$_{\sun}~\text{yr}^{-1}~\text{kpc}^{-2}$. The sSFR, on the other hand, shows a different picture with the three bright regions [region A and the two bar-ends (B and D)] showing similar values meaning equally high efficiency in forming stars, while two other bright regions are revealed, one at the end of the right-hand side spiral arm and one at the left-hand side spiral arm (just next to region B). The nucleus shows a deficiency in sSFR. This is to be expected since the large stellar mass concentrated in the center of the galaxy (top-right panel) indicates that, although there is some current SFR (bottom-left panel) the efficiency is very small.

\subsubsection{Regional variations in ISM structure and SFR}\label{ssubsec:ism}

In order to allow for a more direct comparison of the ISM between the different regions in the galaxy and its properties, we plot the various components discussed above in the bar plot in Fig.~\ref{fig:compont_barplt}. In this plot, the median values of each parameter (surface density of sSFR, SFR, dust mass, \ion{H}{I} mass, H$_2$ mass, and stellar mass) are plotted as bars (from the left to the right) for regions A to G. Since the dynamical range of the different parameters is very different, only the values of the mass surface densities (dust, \ion{H}{I}, H$_2$, and stellar) are indicated by the values in the y-axis (in units of  M$_{\odot}$~pc$^{-2}$). The surface densities of SFR and sSFR are scaled by 10 and by 10$^{10}$, and are in units of M$_{\odot}$~yr$^{-1}$~kpc$^{-2}$ and yr$^{-1}$, respectively.

However, a closer inspection reveals notable distinctions among the regions.
From this comparison, it is evident that it is the nucleus that has the largest stellar mass surface density, followed by the galactic bar region and the bar-end regions B and D with slightly lower values. Region A, the large isolated star-forming region in the southern part of the galaxy, comes with even lower stellar mass density, similar to the values found in the spiral arms (regions F, and G). In terms of molecular hydrogen reservoir, it is the nucleus that has the highest content, followed by the bar-end regions (B, and D) and the bar (region E). The spiral arms (regions F, and G) are less dense environments in terms of molecular gas, with the isolated star-forming region A showing significantly less H$_2$ content. Concerning \ion{H}{I} it is regions A, and D that show the largest density, with the nucleus (region C) being the most deficient one. The bar-end region B, as well as the spiral arms, show similar column densities of \ion{H}{I}, with the bar, itself, slightly less dense. The part of the galaxy that has higher dust surface density is the bar-end region D, followed by the southern bar-end (region B) and by the isolated star-forming region A. Between the two spiral arms, it is the western one (F) which has a larger dust density compared to spiral arm G, with the galactic bar (region E) showing an intermediate dust surface density. The lowest dust density is found in the nucleus (region C).  

Currently, region B (the southern bar-end) is more actively forming stars throughout the galaxy, followed by the other bar-end (region D) while the regions with lower SFR density are the two spiral arms (with F having the lowest SFR density throughout the galaxy). The nucleus of the galaxy, C, and the stellar bar E, have high SFR activity with relatively large SFR surface densities, while the isolated star-forming region A follows at lower rates. The most efficient region in forming stars is region A, followed by the bar-end regions B and D. The less efficient region throughout the galaxy in forming stars is the nucleus (region C) and the bar (region E). The spiral arms have a relatively low sSFR (lower than the bar-ends and the star-forming region A) with the eastern spiral arm, G, being the more efficient among the two.

Summarizing, it is region A which, although it is relatively low in molecular gas density, is the most efficient star-forming region in the galaxy. The two bar-ends (regions B and D) show a large content in molecular gas mass and are the regions that are more actively forming stars. In the nucleus, there is still current star formation taking place at a relatively large rate, but it is the less efficient region to form stars with the lowest molecular gas density. 
This may, in part, reflect the presence of an AGN in this region contributing to the observed emission, though our SFR estimation does not account for its influence. 
Furthermore, a localized enhancement in star formation may have been triggered by the tidal interaction, as proposed by \citet{2022MNRAS.510.3899I}, whose simulations of bar formation in NGC~3627 show a burst-like phase driven by tidal forces. 
A similar picture, although a bit more efficient in forming stars, is the bar region (E). 
The two spiral arms show moderate star-forming activity, with F slightly more active compared to G.

\subsection{The Kennicutt-Schmidt relation}\label{subsec:ks_relation}

Gas is one of the most important ingredients of the ISM with a predominant role in star formation. Between the two
gas phases, molecular and atomic, it is the molecular that
is linked to star formation, with the atomic gas being more
inactive to star formation (acting as a reservoir for the formation of molecular gas). In Fig.~\ref{fig:ken-schm} we plot the surface density of SFR with the surface density of the two different gas components (the atomic, and the molecular gas, left and middle panels, respectively) as well as the total gas (right panel). In all three panels, three diagonal lines indicate the efficiency of the galaxy in consuming the gas. The upper line indicates fast consumption time-scales (0.1~Gyr) assuming that the current SFR keeps constant with time. Similarly, the other two lines indicate gas consumption time-scales of 1, and 10~Gyr, respectively.

\begin{figure}
    \centering
    \captionsetup{labelfont=bf}    
    \includegraphics[width=0.5\textwidth]{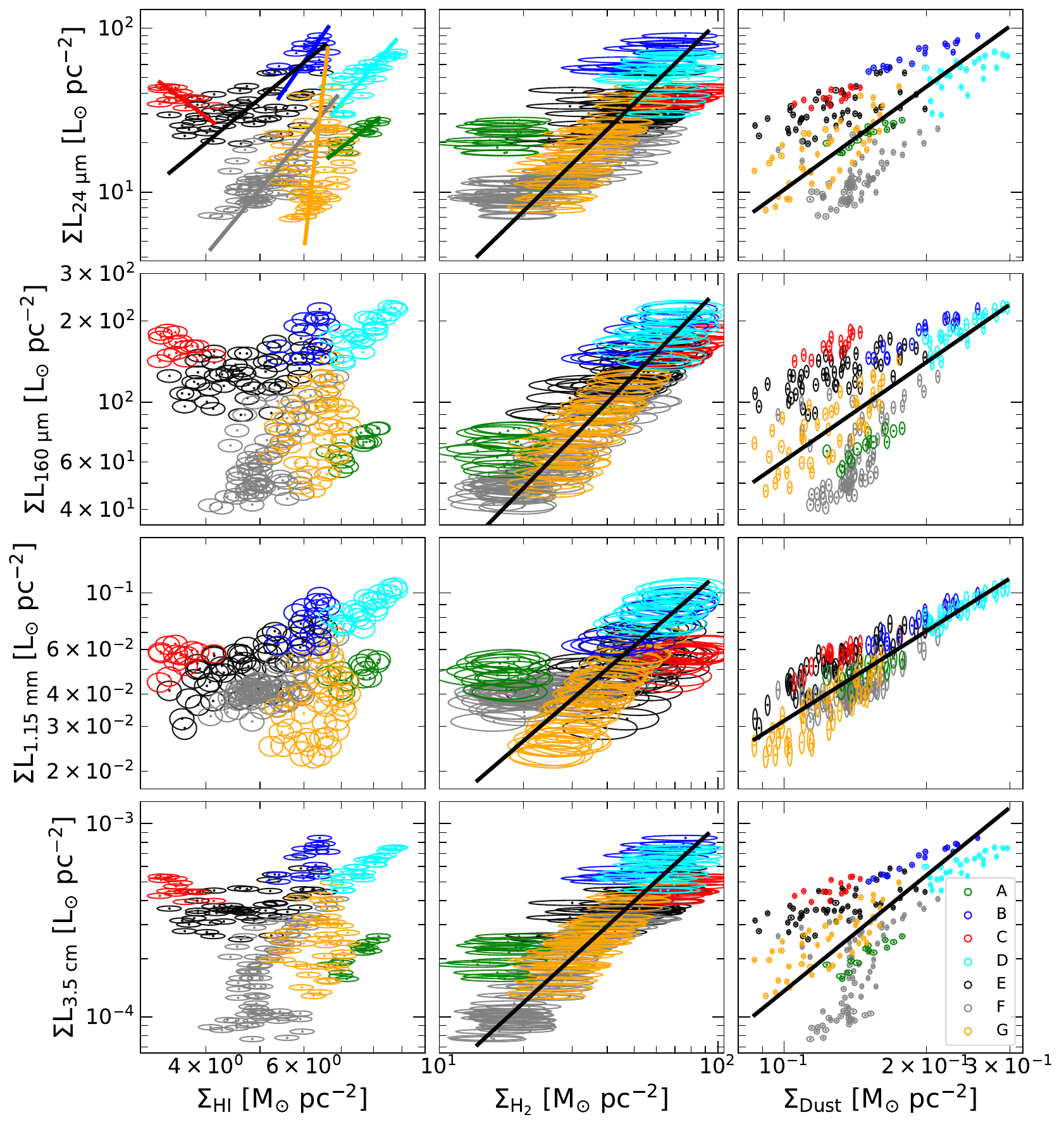} 
    \caption{Relations between the surface brightness in different bands (MIPS~-~24~$\mu$m, PACS~-~160~$\mu$m, NIKA2~-~1.15~mm and VLA~-~3.5~cm; top to bottom) and the surface densities of atomic mass, molecular mass and dust mass surface densities (left to right, respectively).
    Gas surface densities are traced by \ion{H}{I} and CO~(1-0) line emission and interstellar dust is constrained from the SED fitting. 
    The data points represent values of the regions A to G in the galaxy, with the same pixel scale and color coding as in Fig.~\ref{fig:ken-schm}. For the atomic gas mass surface density plots the fitted trends, in each region, are only indicated in the top plot (given that similar behavior is seen in the remaining plots). The fitted correlations for the molecular gas mass and the dust mass surface densities are indicated with black solid lines in the plots with the computed slopes and Pearson's correlation coefficients presented in Table~\ref{tab:fit}. 
    }\label{fig:correlations}
\end{figure}

\begin{table}
\begin{center}
\captionsetup{labelfont=bf} 
\caption{
Parameters for the correlations presented in Fig.~\ref{fig:correlations}.}
\label{tab:fit}
\begin{tabular}{l|cc|cc}
\hline \hline
       & \multicolumn{2}{c|}{$\Sigma_\mathrm{H_2}$}  & \multicolumn{2}{c}{$\Sigma_\mathrm{Dust}$}  \\ \hline
        & $n$        & $r$       & $n$                    & $r$                   \\ \hline
$\Sigma L_\mathrm{24~\mu m}$    & 1.66 $\pm$ 0.08     & 0.785  & 2.09 $\pm$ 0.01     & 0.695  \\
$\Sigma L_\mathrm{160~\mu m}$   & 1.06 $\pm$ 0.04  & 0.926 & 1.21 $\pm$ 0.01   & 0.679   \\
$\Sigma L_\mathrm{1.15~mm}$   & 0.94 $\pm$ 0.05    & 0.761   & 1.17 $\pm$ 0.01   & 0.891  \\
$\Sigma L_\mathrm{3.5~cm}$     & 1.21 $\pm$ 0.06      & 0.882    & 1.99 $\pm$ 0.01    & 0.726  \\
 \hline
\end{tabular}
\end{center}
\begin{minipage}{\linewidth}
\small
\textbf{Notes.} The correlations have the form of $\Sigma L_\mathrm{band} \propto \Sigma^n_\mathrm{(H_2, Dust)}$ with the slopes, $n$, and the Pearson's correlation coefficients, $r$, obtained through least squares fits. 
The correlations have the form of $\Sigma L_\mathrm{band} \propto \Sigma^n_\mathrm{(H_2, Dust)}$ with the slopes, $n$, and the Pearson's correlation coefficients, $r$, obtained through least squares fits. 
\end{minipage}
\end{table}

\begin{figure*}
    \centering
    \captionsetup{labelfont=bf}
    \includegraphics[width=\textwidth]{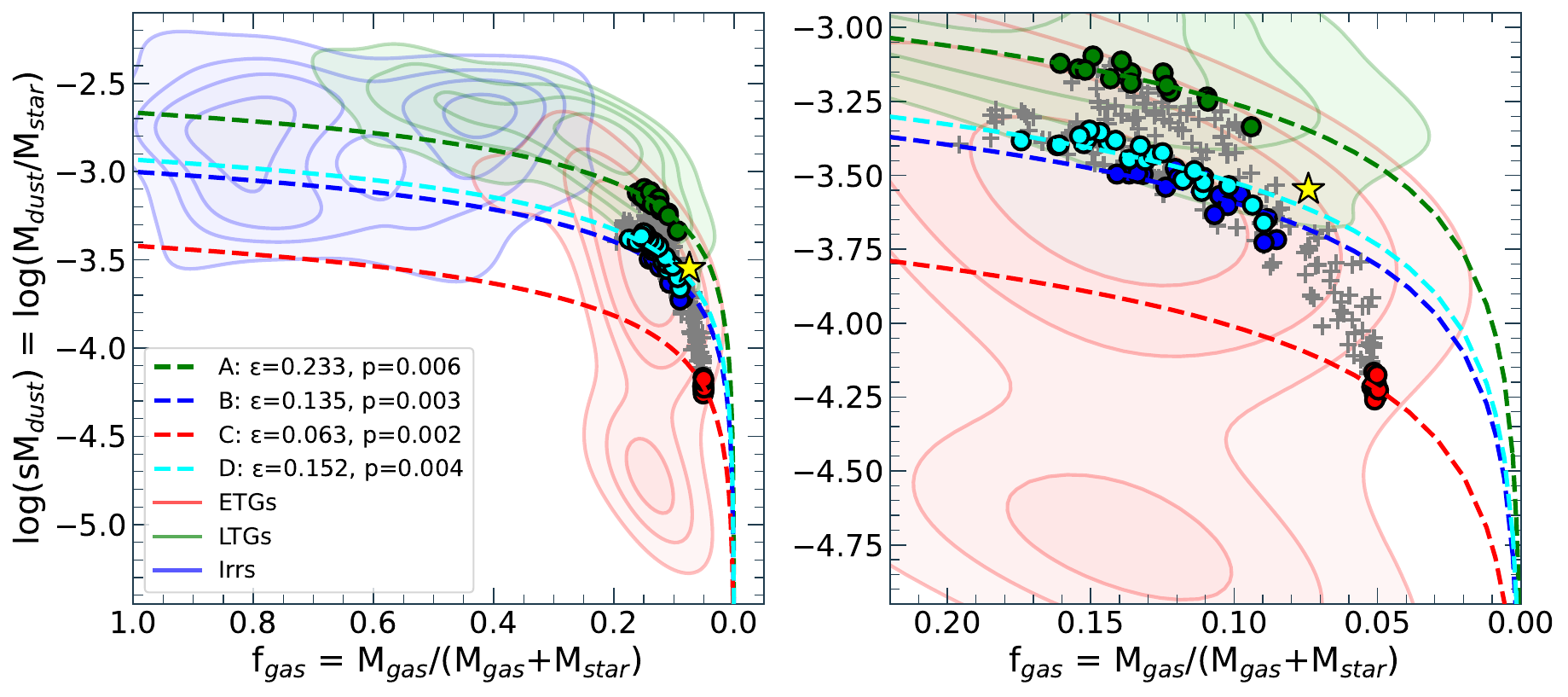}
    \caption{
The gas fraction, $f_\mathrm{gas}$, plotted against the specific dust mass, $sM_\mathrm{dust}$, for NGC~3627. The yellow star indicates the global values found for the galaxy, while the green, blue, red, and cyan circles indicate the pixel values in regions A, B, C, and D, respectively, with the remaining regions of interest indicated with gray crosses. 
The dashed curves are the best fitted `closed-box' models passing through these four regions (see Sect.~\ref{sec:evolution} for a detailed description of the model and the derivation of the relevant parameters). The contours in the background in red, green, and blue represent the kernel density distributions for early-type, late-type, and irregular galaxies, respectively \citep[data from][]{2021A&A...649A..18G}. The right plot is a zoomed-in version of the left plot around the parameter space of the NGC~3627 values.
    }
    \label{fig:dust_evolution}
\end{figure*}

The atomic gas (left panel) shows a variety of gas consumption efficiencies spanning the full range of consumption time-scales ($\sim 0.1 - 10$~Gyr) bracketed by the lines. Most of the areas in the galaxy show a moderate to low efficiency (with time-scales between $\sim 1 - 10$~Gyr) while the active star-forming regions B, and D, as well as the nucleus and the bar show larger consumption time-scales closer to 0.1~Gyr. This might indicate that the atomic gas will very rapidly be transformed to molecular gas in these regions and keep the star formation going on for a longer time. It is worth noticing that the isolated region A shows a lower consumption efficiency of atomic gas compared to the bar-ends (regions B, and D). In this plot, a vertical dashed line marks the saturation threshold of $\Sigma_\mathrm{HI, saturation}=9~\text{M}_{\odot}$~\text{pc}$^{-2}$, as indicated by \cite{2008AJ....136.2846B}, suggesting that NGC~3627 is consistent with this finding. 
We note that our assumptions for the IMF (Kroupa-type) and the CO-to-H$_2$ conversion factor X$_{\mathrm{CO}}$ are consistent with those adopted by \cite{2008AJ....136.2846B}.

The molecular gas (middle panel) shows a tighter correlation with the SFR. The galaxy's efficiency in consuming the molecular gas and transforming it into stars (assuming a constant SFR) is between time-scales of 1 and 10~Gyr. The active star-forming regions, A, B, and D, are on the high end of consumption time-scales ($\sim 1$~Gyr) while the rest of the galaxy (spiral arms, as well as the nucleus, and the bar) need more time (closer to $\sim 10$~Gyr) to transform molecular gas into stars. Comparing to other galaxies \citep[see Fig. 4 in][]{2008AJ....136.2846B} we see that NGC~3627 is consistent with the rest of the sample with a slight offset in the molecular gas consumption towards higher consumption time-scales.

The total gas (right panel) is mostly driven by the behavior of the molecular gas, which is the most dominant gas component (although the combination of the two gas components results in slightly higher consumption time, closer to 10~Gyr). Compared to the \cite{2008AJ....136.2846B} sample (see their Fig. 4), NGC~3627 is, generally, on the higher end of consumption time-scales, indicating that slower processes of transforming gas into stars are taking place within this galaxy.

\subsection{ISM-related scaling relations}

Scaling relations of the ISM with specific bands are very useful tools in identifying those wavelengths that have tighter correlations with specific bands and thus trace specific ISM components, but not only that. Such correlations may reveal differences in the ISM content and unexpected behaviors occurring in specific regions inside the galaxy. Furthermore, the change of the slope of these correlations with wavelength is indicative of the importance of each band in relation to the ISM components. 

We perform such analysis in Fig.~\ref{fig:correlations} with the surface densities of the atomic gas, the molecular gas, and the dust masses (left to right) plotted against indicative bands in the MIR (24~$\mu$m), the FIR (160~$\mu$m), the millimeter (1.15~mm), and the radio (3.5~cm), top to bottom. Here we focus on the brighter regions in the galaxy, which are the areas included in regions A to G (the colors in each panel indicate different regions as described in the inset of the top-left panel).

The atomic gas mass surface density (first column) shows that there is not any uniform and clear correlation among the different regions with the individual bands. On the contrary, each region shows different trends in each band. The most striking difference is a clear anti-correlation in the nucleus (region C) and an almost constant \ion{H}{I} mass surface density for the right spiral arm (region G). All other regions seem to have similar correlations, showing a positive increase in the atomic gas mass surface density with the different wavelengths but all occupying different regions in the plot. The above trends are indicated by lines of least squares fits in each region in the top-left panel (similar trends are present for the rest of the wavelengths).

 \begin{figure*}
    \centering
    \captionsetup{labelfont=bf}
    \includegraphics[width=\textwidth]{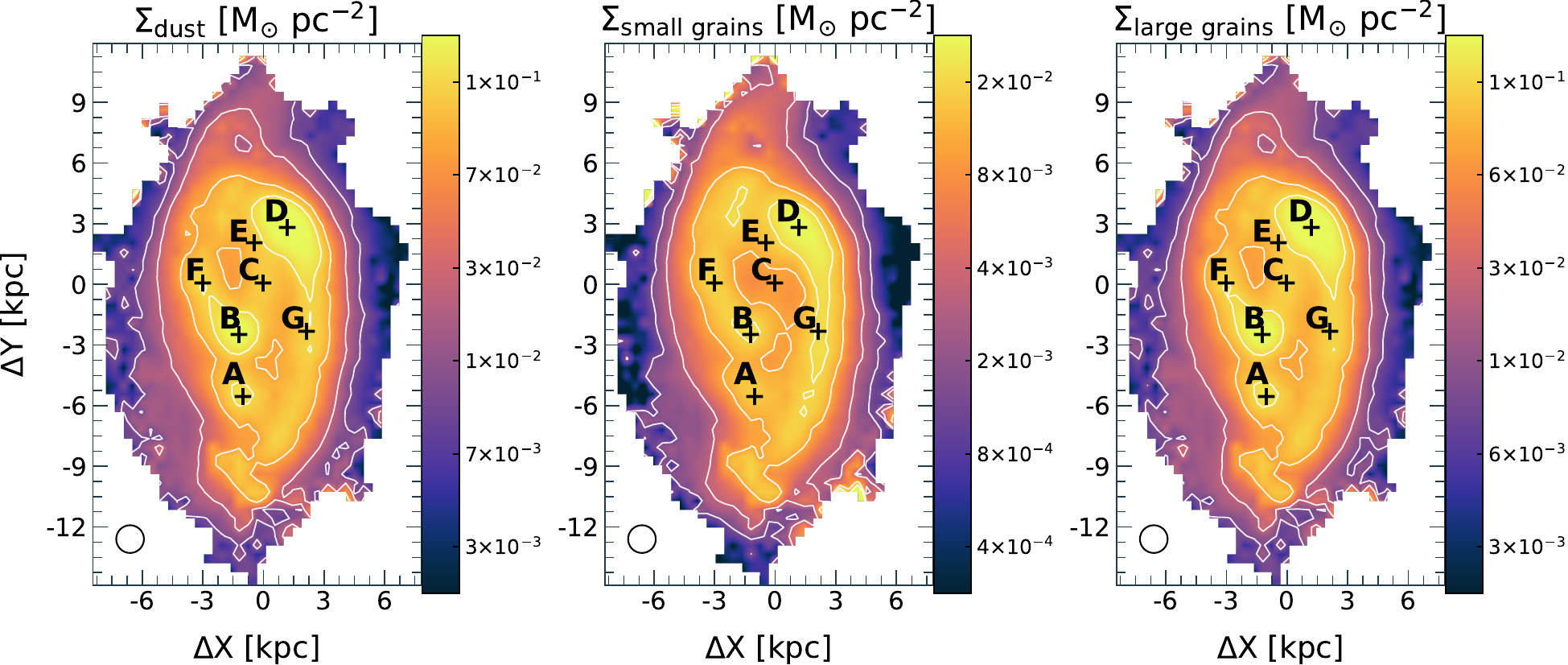}
    \caption{Surface density maps of the total dust mass (left panel), the small grain dust mass (middle panel), and the large grain dust mass (right panel) in NGC~3627 (see Sect.~\ref{subsec:dust_grains} for a detailed description of the derivation of the different dust components).
The beam size is indicated in the bottom-left corner of each map, while distances are in kpc with the galaxy centered at zero. The regions of interest are indicated on each map with white contours guiding the eye on the basic morphological features. 
    }
    \label{fig:dust_grains}
\end{figure*}

The molecular gas mass surface density (middle column) shows a tighter overall correlation with most bands, except for the millimeter wavelengths. This is evident by the power-law functions fitted to the data with their slopes and Pearson's coefficients presented in Table~\ref{tab:fit}. Given the Pearson's coefficient $r$, the band that best traces molecular gas is in the far-IR range (in our case the 160~$\mu$m) with the other wavelengths showing larger scatter. The tight correlation with 160~$\mu$m  reflects the Kennicutt–Schmidt relation (see Sect.~\ref{subsec:ks_relation}), highlighting the link between molecular gas as the star formation fuel and the FIR emission tracing star formation activity. The 160~$\mu$m correlation is also the one with a slope closer to unity, indicating a better correlation between the two quantities. The most uncorrelated case is the NIKA2 band, in which two regions (A and C) are obvious outliers. These regions are deviating from the general trend also in the other wavelengths, but it is in the NIKA2 band that this is really seen. This behavior would imply that region A is missing \ion{H}{I} emission (with respect to the rest of the galaxy) while the nucleus (region C) has \ion{H}{I} content which is more than expected. 

The dust mass surface density (last column) indicates a correlation between the different bands considered here, although with a large scatter in most cases. The band with the largest Pearson's coefficient, $r$, is the NIKA2 band, indicating that these wavelengths are more reliable tracers of the global dust content of the galaxy. This is also supported by the fact that the slope of the power law that is fitted is closer to one compared to the rest of the bands considered here. Physically, this is expected because the 1.15~mm emission is dominated by large grains, which constitute the total dust mass budget.
However, because the dust mass is derived from SED fitting that includes these bands, we tested the trend’s robustness by deriving dust masses from a fit that excludes the 1.15~mm data. The resulting correlation remains similar, consistent with our adopted dust model, in which mm emission primarily traces the cold, large-grain component.

The study by \cite{2008A&A...485..679P} examined the correlation between 20~cm radio continuum, CO~(1-0) emission, and 8~$\mu$m emission in NGC~3627. By fitting data within an 18$^{\prime\prime}$ aperture (see their Fig.~6), they found a slope of 1.12 and a Pearson correlation coefficient of 0.81 for the relationship between 20~cm emission and CO~(1-0) emission. Although our analysis focuses on a shorter wavelength (3.5~cm), we find a similar correlation between radio continuum emission and H$_2$ surface density, as traced by CO~(1-0) emission, with a computed slope $n=1.21$ and a correlation coefficient $r=0.882$.

\subsection{Simplified intra-galactic ISM evolution considerations}\label{sec:evolution}

Relations between the different ISM components may reveal how the different areas throughout the galaxy have evolved till their current stage. Comparing quantities like the gas mass, the dust mass, and the stellar mass has proven valuable proxies for the evolutionary stages of galaxies. Such a study has been conducted by \cite{2021A&A...649A..18G} where the DustPedia and DGS samples of nearby galaxies (of different morphological types) were analyzed using the \texttt{HerBIE} code and available gas mass and SFR data to infer those scaling relations. In this study, galaxies were grouped into three morphological classes (early-type, late-type, and irregular galaxies) and their ISM parameter space was explored (among others) in terms of combinations of the aforementioned parameters \citep[see, e.g., Fig.~8 in][and the various definitions of the parameters in the text]{2021A&A...649A..18G}.

In Fig.~\ref{fig:dust_evolution} we plot such a correlation between the gas fraction ($f_\mathrm{gas}=\frac{M_\mathrm{gas}}{M_\mathrm{gas}+M_\mathrm{star}}$) and the specific dust mass ($sM_\mathrm{dust}=\frac{M_\mathrm{dust}}{M_\mathrm{star}}$). In this plot, the data presented in \cite{2021A&A...649A..18G} are represented as density contours for the early-type galaxies (red contours), the late-type galaxies (green contours), and the irregular galaxies (blue contours). The left panel shows the specific correlation in its full scale, while the right panel shows a zoomed-in area of the parameter space which the data for NGC~3627 occupy.

These plots can be considered as evolutionary tracks of the ISM given that the different morphological types of galaxies host different amounts of ISM in different environments. In \cite{2012A&A...540A..52C} a slightly different sample of galaxies was compared with the predictions of a simple `closed box' model of dust formation and evolution presented in detail in \cite{2001MNRAS.328..223E}. This study showed that the observations of galaxies are consistent with a simple picture in which the amount of dust in galaxies is regulated by the star formation activity and the cold gas content.  

In Fig.~\ref{fig:dust_evolution} we have overplotted the data for NGC~3627 (in circles and stars) as well as the global values indicated by a yellow star. Since we want to focus on the properties in regions of the galaxy that can be well-defined and could be considered isolated from the rest of the body of the galaxy, we highlight the star-forming regions A, B, and D, as well as the nucleus of the galaxy (region C) with colors as indicated in the inset of the left panel, with the remainder of the galaxy indicated by gray crosses. In a very simplistic approach, one could consider these regions isolated and approximated as `closed boxes' inside the galaxy. 

This plot shows that this galaxy, if considered as a whole (yellow star), falls in the transition range between late-type and early-type galaxies (although this can only be considered as a qualitative argument since the extent indicated by the contours is relatively high). It is, nevertheless, clear from this plot that inactive regions (in terms of star formation) are located in the parameter space of the early-type galaxies, while active star-forming regions, such as A, B, and D, occupy the parameter space where early-type and late-type galaxies overlap.

To better understand the properties of these distinct regions inside the galaxy we made use of the `closed-box' model approach described in \cite{2001MNRAS.328..223E}. As in \cite{2012A&A...540A..52C}, we have used Eqs.~(5) and (11) in \cite{2001MNRAS.328..223E} to determine how the specific dust mass evolves with the gas fraction. To keep the investigation more robust, we have held four parameters constant—these parameters are widely regarded as accurate for such models—and we have varied the two other parameters, which appear to have a greater impact on the model's outcomes. The ones that are kept fixed are the efficiency of dust condensation from heavy elements made in stellar winds from massive stars and supernovae \citep[$\chi_1=0.2$;][]{2003MNRAS.343..427M} and from asymptotic giant branch stars \citep[$\chi_2=0.5$;][]{2008A&A...479..453Z}, the fraction of mass of the ISM locked into stars \citep[$a=0.8$; consistent with a][IMF]{2003PASP..115..763C}, and the fraction of dust destroyed by star formation \citep[$\delta=0.3$;][]{2001MNRAS.328..223E}. The parameters that were varied (using a $\chi^2$ approach) are the effective yield ($p$; the ratio of the mass of metals produced by stars which remain in the ISM, relative to the total gas mass) and the fraction of the ISM where mantles can grow ($\epsilon$). Dashed lines, matching the colors of the symbols for regions A, B, C, and D, illustrate the evolutionary paths based on the observations of these four distinct regions of interest.
These regions were chosen because they represent a range of evolutionary stages within NGC~3627, covering both extreme and intermediate cases.

For a more clear visualization of the model and observations comparison, we present the zoomed-in region of this plot in the right panel of Fig.~\ref{fig:dust_evolution}. The optimal values of $\epsilon$ and $p$ that align with the data are displayed in the inset on the left panel plot. Notably, less active areas, such as the galaxy's nucleus C, follow evolutionary tracks with low $\epsilon$ and $p$ values, whereas these values rise for regions B, D, and A. Region A, the isolated \ion{H}{II} region, exhibits $\epsilon$ and $p$ values approximately twice as high as those of B and D (the two bar-ends) and more than threefold compared to the nucleus. 
The elevated $\epsilon$ in region A might be attributed to the X-ray hot gas identified in NGC~3627 (see Fig.~\ref{fig:2mm}). 
Given that mantles can be effectively photodissociated in environments with intense UV and X-ray radiation, it becomes apparent from Fig.~\ref{fig:2mm} that the nucleus, which exhibits the most substantial X-ray radiation, along with the bar-ends, is expected to show a decreased proportion of the ISM where mantles can flourish compared to region A, which seems to lack X-ray radiation. The larger effective yield $p$ observed in region A could suggest that, considering its isolation relative to the rest of the galaxy, the metals generated by stars remain confined to this region with minimal losses over time. Conversely, the two bar-ends and the nucleus, linked by the galactic bar and exhibiting greater dynamical instabilities, could impact the metal content in these areas. Galactic outflows in these regions might also serve as additional factors leading to reduced metal abundances.

\subsection{Distribution of small and large dust grains}\label{subsec:dust_grains}

Following the analysis of \cite{2023A&A...679A...7K}, we categorize dust grains into two populations, the small grains (with sizes smaller than 15~\AA), mostly being responsible for aromatic features seen in the MIR spectra, and the large grains which are the sources responsible for the bulk of the radiation emitted at FIR and submm wavelengths. This has been made possible through the usage of the \texttt{THEMIS} dust model, implemented in \texttt{HerBIE}, which, based on laboratory data, accounts for both aromatic and aliphatic mid-IR features. In the \texttt{THEMIS} model, the small, partially hydrogenated amorphous carbon grains [a-C(:H)], which, despite their varying levels of hydrogenation, exhibit similarities to PAHs.
Additionally, it incorporates dust grain evolution processes, including Fe and FeS nano-inclusions. \texttt{HerBIE} further parameterizes these dust grain populations according to their size, distinguishing between very small a-C(:H) (VSAC; <~7~\AA), small a-C(:H) (SAC; $7-15$~\AA), and medium-to-large a-C(:H) (MLAC; >~15~\AA). In this study \citep[as in][]{2023A&A...679A...7K} we refer to VSAC and SAC as small grains, while MLAC represents the large grain population. 
The mass fraction of small grains, denoted as $q_{AF}$, was treated as a free parameter in the model.

\begin{figure}
    \centering
    \captionsetup{labelfont=bf}
    \includegraphics[width=0.47\textwidth]{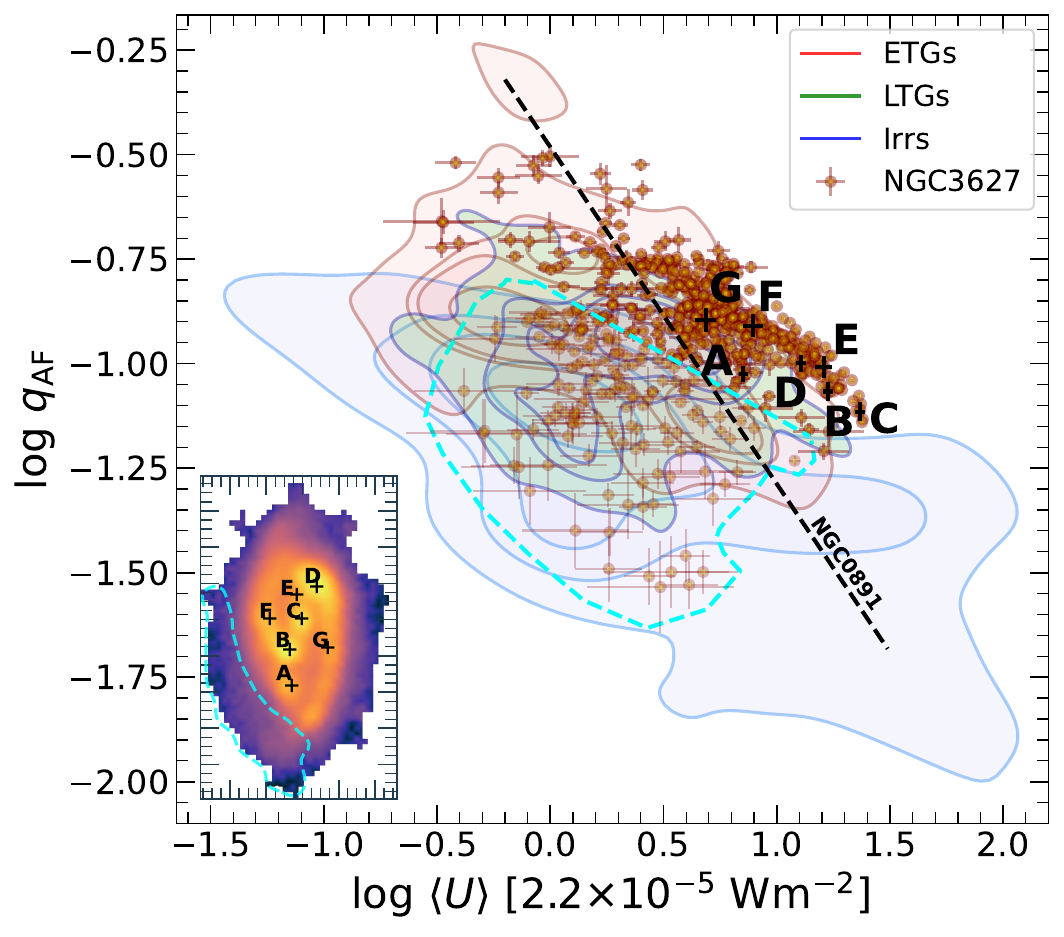}
    \caption{The mass fraction of small dust grains, q$_\mathrm{AF}$, versus the average ISRF, $\langle U\rangle$. The brown points are the pixel values of NGC~3627 along with their associated error-bars. Black letters and crosses indicate the regions of interest considered in this study (see the inset, SPIRE~-~350~$\mu$m~map). The background contours indicate the kernel density distributions for early-type, late-type, and irregular galaxies (red, green, and blue, respectively; see also Fig.~\ref{fig:dust_evolution}), based on the data from \citet{2021A&A...649A..18G}.
    The points included within the cyan-colored dashed contour in the plot are located in the south-west part of the galaxy (see the cyan-colored region of the galaxy in the inset map).
    The black dashed line represents the fit of the data values for the disk of NGC~891, as derived from the findings in \citet{2023A&A...679A...7K}.
    }
    \label{fig:qAF_Uav_G21}
\end{figure}

From the global SED fitting, we derived the dust mass of small grains in NGC~3627 to be $M_\mathrm{small~grains}~=2.72\times10^6$~\text{M}$_{\odot}$, accounting for 13\% of the total dust mass (see Table~\ref{tab:global_sed}). According to Eq.~(10) of \cite{2021A&A...649A..18G}, at 0.85~Z$_{\odot}$, $q_\mathrm{AF}\simeq(15\pm9)$\%. For comparison, the mass fraction of small a-C(:H) in the solar neighborhood derived from the \texttt{THEMIS} model is 17\% \citep{2021A&A...649A..18G} while for NGC~891 it is 9.5\% \citep{2023A&A...679A...7K}. 

It is interesting to further explore how these dust components vary at spatial scales across different galactic regions.
Figure~\ref{fig:dust_grains} illustrates the distribution of the total dust mass (left panel) analyzed into small grains (middle panel) and large grains (right panel). The distributions of small and large grains, although following a similar overall morphology with total dust (see Sect.~\ref{subsec:ISM_morph}), do reveal some differences. The most striking difference is the lack of small dust grain content in the central region of the galaxy (the nucleus and the bar; regions C and E). 
The deficit of small dust grains in the central regions is consistent with the findings of \cite{2025A&A...694A..87K}, who also reported a suppression of [\ion{C}{II}] (158~$\mu$m) emission in these areas. Small grains are considered to play a key role in the photoelectric heating of the ISM. By absorbing UV photons from young stars and releasing electrons, they heat the surrounding gas. This process enhances the emission of [\ion{C}{II}], one of the primary coolants of neutral gas. The reduced abundance of small grains towards the center of the galaxy can therefore lead to less efficient heating and, consequently, a lower [\ion{C}{II}] emission.

Furthermore, the presence of a low-luminosity AGN in the galaxy's center provides an additional explanation for the [\ion{C}{II}] deficit. \cite{2017ApJ...834....5S} found that galaxies hosting such AGN often exhibit significantly reduced [\ion{C}{II}] emission, with deficits increasing sharply toward the nucleus. This aligns with our observations and suggests that both dust grain properties and AGN activity influence the thermal balance of the neutral gas \citep[see, e.g.,][]{2001ApJ...548L..73H}.
The morphology of the distribution of the small dust grains can be considered to follow a ring-like structure similar to that of the atomic gas (see Fig.~\ref{fig:components}) while large dust grains follow the distribution of the total dust mass (as expected since this is the dominant contributor of the dust material in the galaxy). Another difference between the two maps is that the small dust grains show a flatter distribution throughout the galaxy, with the bright regions in the galaxy (e.g., regions A, B, C, D, E) being very pronounced in the large grain distribution only. These regions account for 24\% of the total large dust grain content, while only for 17\% of the total small dust grain material.

\subsubsection{The role of ISRF in shaping small dust grain abundances}\label{ssubsec:qaf_Uav}

To further investigate in which environments these two populations of dust grains reside, we plot in Fig.~\ref{fig:qAF_Uav_G21} the mass fraction of small grains ($q_\mathrm{AF}$), against the mean intensity of the ISRF ($\langle U\rangle$). Both of these quantities are free-varying parameters of the SED fitting process. In this figure, the values derived for NGC~3627 (orange points) are presented. In addition to the SED modeled parameters for the two galaxies, the background contour regions in this plot represent the kernel density distributions of galaxies classified as early-type, late-type, and irregular \citep[same illustration as in Fig.~\ref{fig:dust_evolution}; see also][]{2021A&A...649A..18G}.

The observed trend is an anti-correlation with the higher the ISRF the lower the small dust grain content, which comes naturally considering the fact that small dust grains are much more efficiently destroyed in harsh radiation environments \citep[see, e.g.,][]{2024A&A...685A..76E}. This general trend is seen in both galaxies, NGC~3627 and NGC~891 \citep[see Fig. 9 in][for the latter]{2023A&A...679A...7K}, but also when galaxies of different morphologies are considered \citep[see the background density contours, and also,][]{2007ApJ...663..866D, 2008ApJ...672..214G, 2013MNRAS.431.2006K, 2015A&A...582A.121R, 2021A&A...649A..18G}.
The parameter space encompassing the values for NGC~3627 (but also NGC~891) aligns well with the background density contours, mostly in the regions where early- and late-type galaxies reside. From this analysis conducted for the global galaxy properties, it is evident that ETGs show a slightly larger fraction of small dust grains compared to LTGs (although they span similar ranges in ISRF), while it is irregular galaxies that show a lower fraction of small dust grains with increasing ISRF (exhibiting larger ISRF compared to the other two types of galaxies). A power-law correlation between the two quantities gives a slope of $-0.25$. This is very similar to the slope found for the disk of NGC~3627, but it is quite different from that of NGC~891 \citep[fitting the data in Fig.~9 in][we find a much steeper slope of $-0.80$]{2023A&A...679A...7K}. We speculate that this is due to inclination effects, with the quantities derived for NGC~891 being vastly affected by the fact that the properties in different environments are mixed along the line-of-sight resulting in underestimation, mainly, of the fraction of small dust grains. Unlike NGC~3627 where distinct regions like the nucleus or the bar-ends or \ion{H}{II} regions are separated from the rest of the galaxy and thus measure their intrinsic properties, in NGC~891 such regions may be mixed with other such regions, but also with the spiral arm, inter-arm regions and thus result in underestimation of $q_\mathrm{AF}$. 

The position of the regions of interest in this plot is very similar to those in NGC~891 \citep[see Fig. 9 in][]{2023A&A...679A...7K} with the nucleus of the galaxy experiencing the highest $\langle U\rangle$ followed by the bar and the two bar-ends (B and D) and the isolated \ion{H}{II} region A with lower ISRF and higher small grain fractions. The spiral arms (F and G) show, on average, even lower $\langle U\rangle$ values and higher $q_\mathrm{AF}$. It is interesting to notice that in NGC~891 \citep{2023A&A...679A...7K} we see the parameter space occupied by both the disk and the halo of the galaxy (which is made possible due to the edge-on configuration of the galaxy), which is not evident in NGC~3627. What is evident, though, in NGC~3627, is a distinct area of the galaxy with values of $q_\mathrm{AF}$ below $\sim10$\% and log$\langle U\rangle$ between $-0.5$ and 0.8 (indicated by the cyan dashed contour). This region with low $q_\mathrm{AF}$ values is in the southern part of the galaxy (see the inset map in Fig.~\ref{fig:qAF_Uav_G21}). This part happens to occupy the parameter space extended even to irregular galaxies (with a large fraction of them considered to be the result of tidal interactions, mergers, or external perturbations). We speculate that this is considered to be the result of the tidal interactions with the neighboring galaxy NGC~3628 in the Leo Triplet Group. The asymmetry that is seen in the galaxy (between north and south) supports this argument in the sense that the northern part is more ordered, characteristic of an undistorted galaxy, while the more elongated, southern part of the galaxy reflects the possible interaction with NGC~3628. These dynamical instabilities that occurred in the past could be the cause of the deficit of smaller grains in these regions.

\section{Conclusions}\label{sec:conclu}

Our analysis is based on SED modeling of the dust and radio emission in NGC~3627 at $\sim$~kpc scales, leveraging the high resolution NIKA2 maps at 1.15 and 2~mm, presented, for the first time, in this study. All input maps were pre-processed and homogenized using the advanced \texttt{HIP} pipeline \citep{pantoniprep}. The dust emission properties were constrained by the \texttt{THEMIS} dust grain model \citep{2013A&A...558A..62J, 2017A&A...602A..46J}, implemented within the \texttt{HerBIE} SED fitting code \cite{2021A&A...649A..18G}.
Our main findings can be summarized as follows:

-- The NIKA2 emission of NGC~3627 at 1.15~mm and 2~mm closely resembles the morphology of the FIR and the molecular gas emission, as traced by CO~(1-0), with four prominent bright regions  observed: the nucleus, the two bar-ends, and a large, isolated H II region located approximately 3 kpc south of the southern bar-end of the galaxy.
At 1.15~mm, the emission is primarily dominated by cold dust (considering an 8\% contribution from CO~(2-1) line contamination), while at 2~mm, synchrotron and free-free emission contribute at a level of $\sim10$\% and $\sim5$\%, respectively. The synchrotron emission is strongest at the southern bar-end, where the surrounding X-ray emission is also intense. A possible recent collision with a dwarf galaxy from the eastern side of the galactic disk has been proposed to explain the X-ray emission, and the proximity of the synchrotron peak further supports this scenario. Interestingly, our analysis reveals that free-free emission exhibits two distinct behaviors, which we attribute to the presence of ULXs by influencing the ionized gas flow through shocks.

-- The distribution of different ISM components (dust, atomic, and molecular gas) as well as the stellar content of the galaxy and the SFR and sSFR were computed. The northern bar-end of NGC~3627 is brighter across all ISM components, while the southern bar-end exhibits higher SFR. The lowest current star formation activity is found in the spiral arms, whereas ongoing star formation is present in the nucleus, though at a reduced level. In terms of efficiency, the isolated \ion{H}{II} region is the most efficient, followed by the bar-ends. In contrast, the nucleus and the bar exhibit the lowest star-formation efficiency, suggesting that while star formation is occurring, it is relatively inefficient.

-- The Kennicutt-Schmidt relation in NGC~3627 shows that atomic gas in the brightest regions has rapid consumption time-scales ($\sim0.1$~Gyr), indicating quick conversion to molecular gas. Molecular gas, however, is more tightly correlated with star formation. Active regions (A, B, and D) have longer consumption times ($\sim1$~Gyr), while less active areas, such as the spiral arms, nucleus, and bar, require much longer ($\sim10$~Gyr) to form stars. The total gas consumption time is primarily driven by the molecular gas component, with a slightly higher total time when combining both gas types ($\sim10$~Gyr).

-- The investigation of the correlations between the surface densities of atomic gas, molecular gas, and dust with various infrared, millimeter, and radio bands reveals different trends across regions. The atomic gas mass surface density shows no clear or uniform correlation with the bands, with notable differences between regions, such as an anti-correlation in the nucleus and a constant \ion{H}{I} mass surface density in the right spiral arm. The molecular gas mass surface density shows a stronger correlation, particularly with the 160~$\mu$m band, which has a slope close to unity. The NIKA2 millimeter band, however, shows weaker correlations, with the nucleus and isolated star-forming region as outliers. The dust mass surface density shows a general correlation with all bands, although with significant scatter. The NIKA2 band stands out with the highest Pearson's coefficient, making it a more reliable tracer of the galaxy’s global dust content, given our adopted dust model and fitting approach.

-- Our examination of the relationship between gas fraction and specific dust mass, interpreted using the simple closed-box model of \cite{2001MNRAS.328..223E}, provides insights into the evolutionary state of the ISM. The galaxy as a whole falls in the transition range between late-type and early-type galaxies, while inactive regions align with early-type galaxies, and active star-forming regions occupy an intermediate space. The larger effective yield observed in region A suggests that metals remain confined due to its relative isolation. In contrast, the bar-ends and nucleus, influenced by the galactic bar and associated dynamical instabilities, may experience metal loss through galactic outflows, affecting their chemical composition.

-- Our analysis estimates the total dust mass of the galaxy to be 2~$\times$~10$^7$~M$_{\odot}$, corresponding to a gas-to-dust mass ratio of 370, consistent with low-metallicity conditions. Small grains (SAC and VSAC) account for 13\% of the total dust mass. Interstellar dust is primarily concentrated in the galactic bar, the spiral arms, and a prominent distant star-forming region, with a significant amount also extending into the southern part of the galactic disk, forming a tidal interaction tail within the Leo Triplet Group. A deficit of small grains is observed in the galactic bar, the isolated \ion{H}{II} region, and the southern tidal tail, whereas the northern part of the galaxy exhibits an increased small grain fraction.

\section*{Acknowledgements}

The research work was supported by the Hellenic Foundation for Research and Innovation (HFRI) under the 3rd Call for HFRI PhD Fellowships (Fellowship Number: 5357). We would like to thank the IRAM staff for their support during the campaigns. The NIKA2 dilution cryostat has been designed and built at the Institut N\'eel. In particular, we acknowledge the crucial contribution of the Cryogenics Group, and in particular Gregory Garde, Henri Rodenas, Jean Paul Leggeri, Philippe Camus. 
The NIKA2 data were processed using the Pointing and Imaging In Continuum (PIIC) software, developed by Robert Zylka at the Institut de Radioastronomie Millim\'etrique (IRAM) and distributed by IRAM via the GILDAS pages. PIIC is the extension of the MOPSIC data reduction software to the case of NIKA2 data. 
This work has been partially funded by the Foundation Nanoscience Grenoble and the LabEx FOCUS ANR-11-LABX-0013. This work is supported by the French National Research Agency under the contracts "MKIDS", "NIKA" and ANR-15-CE31-0017 and in the framework of the "Investissements d’avenir" program (ANR-15-IDEX-02). 
This work has benefited from the support of the European Research Council Advanced Grant ORISTARS under the European Union's Seventh Framework Programme (Grant Agreement no. 291294). 

\section*{Data Availability}

The NIKA2 observations underlying this article will be available on the IRAM science data archive\footnote{\label{iram_archive}\href{http://iram-institute.org/science-portal/data-archive/}{http://iram-institute.org/}}. The derived physical parameter maps obtained from SED fitting will be available at zenodo.

\section*{}
\small{$^{1}$National Observatory of Athens, IAASARS, Ioannou Metaxa and Vasileos Pavlou GR-15236, Athens, Greece\\
$^{2}$Department of Astrophysics, Astronomy \& Mechanics, Faculty of Physics, University of Athens, Panepistimiopolis, GR-15784 Zografos, Athens, Greece\\
$^{3}$Universit\'e Paris-Saclay, Universit\'e Paris Cit\'e, CEA, CNRS, AIM, 91191, Gif-sur-Yvette, France\\
$^{4}$Universit\'e C\^ote d'Azur, Observatoire de la C\^ote d'Azur, CNRS, Laboratoire Lagrange, France \\
$^{5}$School of Physics and Astronomy, Cardiff University, Queen's Buildings, The Parade, Cardiff, CF24 3AA, UK \\
$^{6}$Institut de Radioastronomie Millim\'etrique (IRAM), Avenida Divina Pastora 7, Local 20, E-18012, Granada, Spain \\
$^{7}$Sterrenkundig Observatorium Universiteit Gent, Krijgslaan 281 S9, B-9000 Gent, Belgium \\
$^{8}$Aix Marseille Univ, CNRS, CNES, LAM (Laboratoire d'Astrophysique de Marseille), Marseille, France \\
$^{9}$Institut N\'eel, CNRS, Universit\'e Grenoble Alpes, France \\
$^{10}$Institut de RadioAstronomie Millim\'etrique (IRAM), Grenoble, France \\
$^{11}$Univ. Grenoble Alpes, CNRS, Grenoble INP, LPSC-IN2P3, 53, avenue des Martyrs, 38000 Grenoble, France \\
$^{12}$Dipartimento di Fisica, Sapienza Universit\`a di Roma, Piazzale Aldo Moro 5, I-00185 Roma, Italy \\
$^{13}$Univ. Grenoble Alpes, CNRS, IPAG, 38000 Grenoble, France \\
$^{14}$Institute for Research in Fundamental Sciences (IPM), School of Astronomy, Tehran, Iran \\
$^{15}$Centro de Astrobiolog\'ia (CSIC-INTA), Torrej\'on de Ardoz, 28850 Madrid, Spain \\
$^{16}$IRAP, Université de Toulouse, CNRS, UPS, IRAP, Toulouse Cedex
4, France\\
$^{17}$Universit\'e Paris-Saclay, CNRS, Institut d'astrophysique spatiale, 91405, Orsay, France \\
$^{18}$High Energy Physics Division, Argonne National Laboratory, 9700 South Cass Avenue, Lemont, IL 60439, USA \\
$^{19}$LUX, Observatoire de Paris, PSL Research University, CNRS, Sorbonne Universit\'e, UPMC, 75014 Paris, France \\ 
$^{20}$Institute of Space Sciences (ICE), CSIC, Campus UAB, Carrer de Can Magrans s/n, E-08193, Barcelona, Spain \\
$^{21}$ICREA, Pg. Llu\~As Companys 23, Barcelona, Spain \\
$^{22}$STAR Institute, Quartier Agora - All\'ee du six Ao\"ut, 19c B-4000 Li\'ege, Belgium \\
$^{23}$Dipartimento di Fisica, Universit\`a di Roma “Tor Vergata”, Via della Ricerca Scientifica 1, I-00133 Roma, Italy	\\
$^{24}$Institute of Astronomy, KU Leuven, Celestijnenlaan 200D, 3001 Leuven, Belgium \\
$^{25}$School of Physics and Astronomy, University of Leeds, Leeds LS2 9JT, UK \\
$^{26}$Laboratoire de Physique de l'\'Ecole Normale Sup\'erieure, ENS, PSL Research University, CNRS, Sorbonne Universit\'e, Universit\'e de Paris, 75005 Paris, France \\
$^{27}$INAF-Osservatorio Astronomico di Cagliari, Via della Scienza 5, 09047 Selargius, IT \\
$^{28}$Institut d'Astrophysique de Paris, CNRS (UMR7095), Sorbonne Universit\'e, 98 bis boulevard Arago, 75014 Paris, France \\
$^{29}$University of Lyon, UCB Lyon 1, CNRS/IN2P3, IP2I, 69622 Villeurbanne, France}



\bibliographystyle{mnras}
\bibliography{references} 



\appendix

\section{Maps of NGC~3627 used and their properties}\label{sec:appendA}

\begin{figure*}
    \centering
    \captionsetup{labelfont=bf}
    \includegraphics[width=0.85\textwidth]{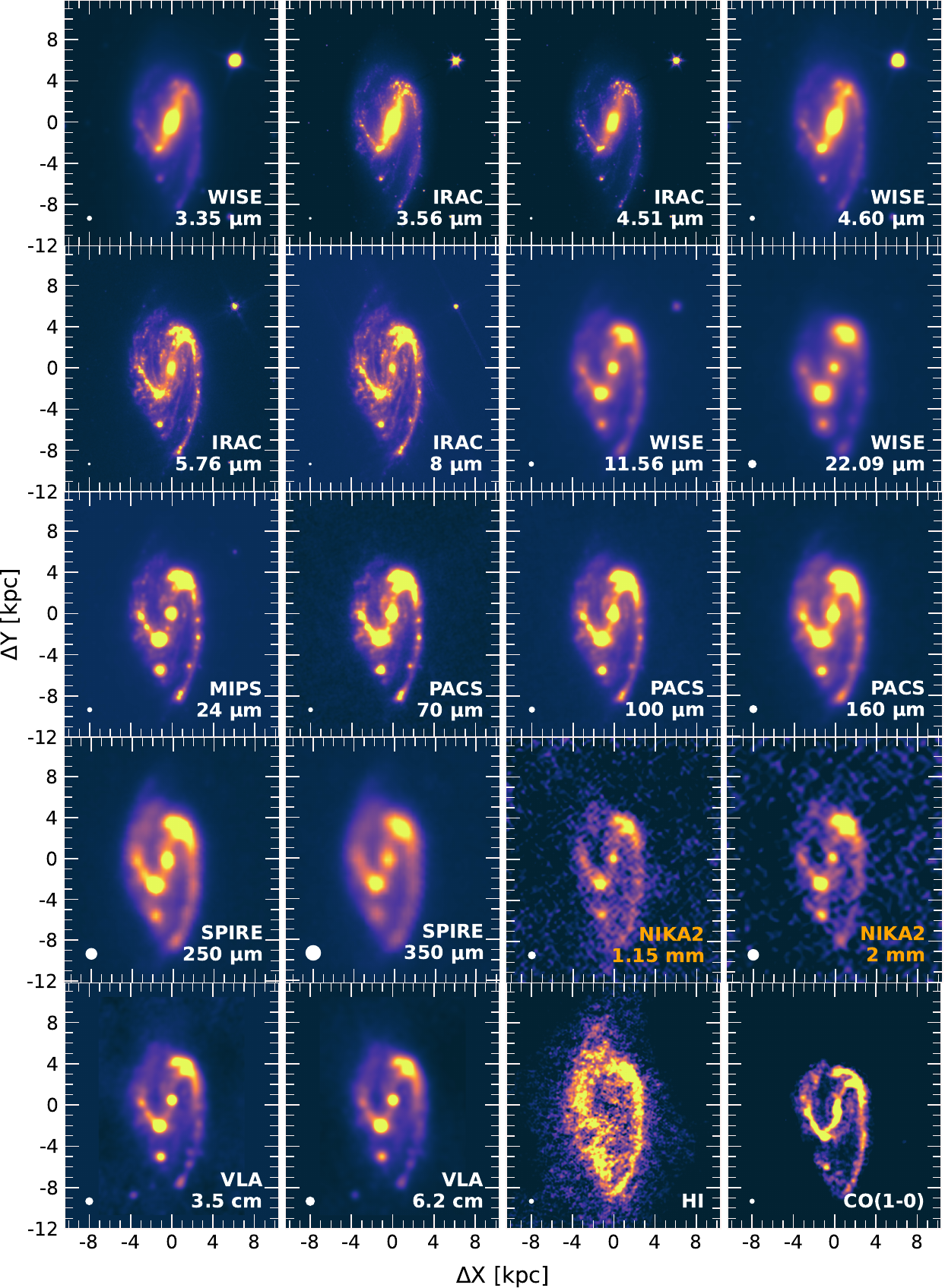}
    \caption{Maps of NGC~3627 used to constrain the spatially resolved SED fitting. The maps (including the new NIKA2 observations presented in this study) span a wavelength range from 3.4~$\mu$m to 6.2~cm (top-left to bottom-right in increasing wavelength) with their native resolution indicated as a white circles in the bottom-left corner of each panel (see their basic properties in Table~\ref{tab:photometry}). The last two panels show the \ion{H}{I} and CO~(1-0) maps that are used as priors in the SED modeling. All maps are centered at RA$_\mathrm{J2000}~= 11^\mathrm{h}20^\mathrm{m}15^\mathrm{s}$, DEC$_\mathrm{J2000}~= +12^{\circ}59^{\prime}30^{\prime\prime}$ and distances are measured in kpc from the galactic center.
    }
    \label{fig:maps}
    \end{figure*}

\begin{figure}
    \centering
    \captionsetup{labelfont=bf}    \includegraphics[width=0.49\textwidth]{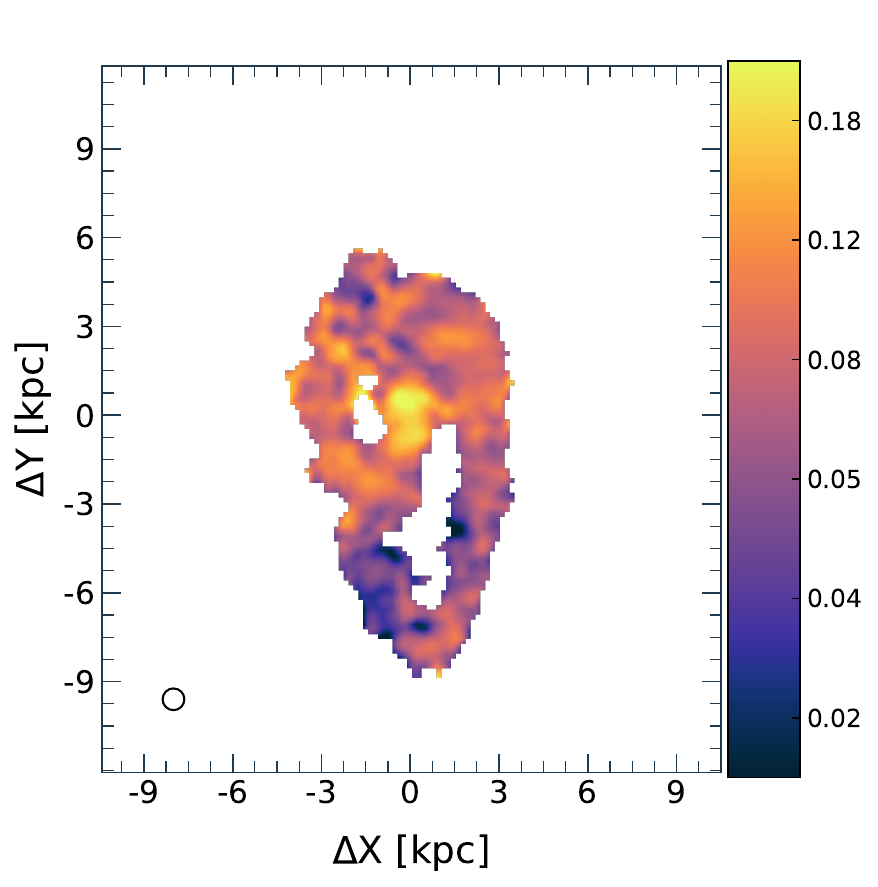}
    \caption{Fraction of the CO~(2-1) contamination of the NIKA2 1.15~mm emission. The map is presented at a resolution of 13.4$^{\prime\prime}$, the resolution of the IRAM/CO~(2-1) map. The subtraction of the line contamination has been made using the \texttt{HIP} pipeline (see Sect.~\ref{sec:process}).}
    \label{fig:co21}
\end{figure}

Fig.~\ref{fig:maps} presents all the maps of NGC~3627 that were used in the resolved SED fitting analysis. These are 18 maps in the wavelength range from 3.4 $\mu$m to 6.2 cm (including the new NIKA2 maps at 1.15~mm and 2~mm observations presented in this study). The \ion{H}{II} and CO~(1-0) maps, also presented in this figure, were used as priors in the hierarchical Bayesian SED modeling tracing the atomic and the molecular gas distribution. The properties of the maps are listed in Table~\ref{tab:photometry} and discussed in detail in Sect.~\ref{sec:data}. 

As discussed in Sect.~\ref{sec:process}, we estimated the level of contamination from the CO~(2–1) line within the NIKA2 1.15~mm band by using available CO~(2–1) data. The resulting contamination map is shown in Fig.~\ref{fig:co21}, expressed as a fraction of the observed NIKA2 1.15~mm spectral emission. 

\section{Resolved SED modeling parameters and goodness of the fit}\label{sec:appendB}

\begin{figure}
    \centering
    \captionsetup{labelfont=bf}    \includegraphics[width=0.49\textwidth]{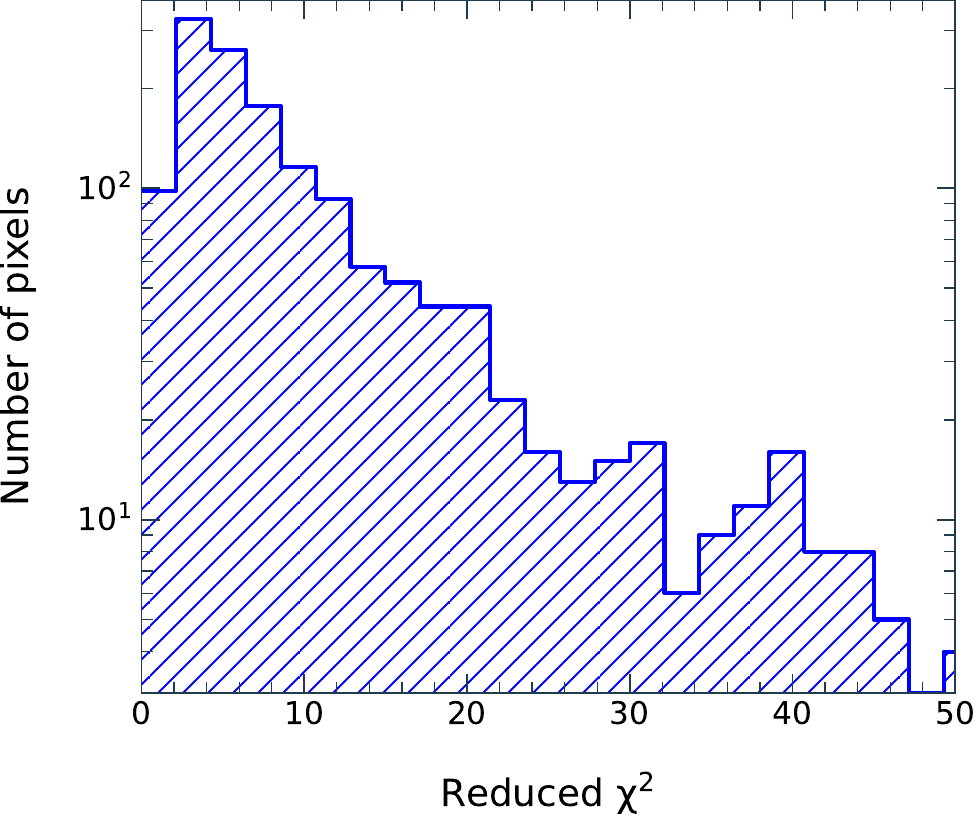}
    \caption{Distribution of the reduced $\chi^2$ values of the spatially resolved fit of NGC~3627, performed with the \texttt{HerBIE} SED fitting code.}
    \label{fig:chi2}
\end{figure}

As detailed in Sect.~\ref{subsec:seds}, 18 maps of NGC~3627 (see Fig.~\ref{fig:maps}, and Table~\ref{tab:photometry}) were used to model, spatially, the SED of the galaxy by varying 10 parameters of the \texttt{HerBIE} code. Following a 3$\sigma$ threshold cut-off of each maps, and after a homogenization process, the maps were all convolved to the same 25$^{\prime\prime}$ resolution and a pixel scale of 8$^{\prime\prime}$, before fed to the \texttt{HerBIE} SED fitting code. The goodness of the fit of the model to the data is presented in Fig.~\ref{fig:chi2} showing the distribution of the reduced $\chi^2$ values. It turns out that 64\% of the 1465 fitted pixels (approximately 940) show reduced-$\chi^2 < 10$, 84\% (approximately 1230 pixels) show reduced-$\chi^2 < 20$, and 92\% (approximately 1350 pixels) show reduced-$\chi^2 < 30$. The peak of the distribution is at reduced-$\chi^2$=3.9. Table~\ref{tab:spatial_sed} lists the parameter values derived from the resolved SED fitting analysis in the regions of interest A to G (see Fig~\ref{fig:seds}). The reported values are averages within these regions. Typical SEDs in these regions are presented in Fig.~\ref{fig:seds}.

\begin{table*}
    \centering
    \captionsetup{labelfont=bf} 
    \caption{Median values of the model parameters of NGC~3627 derived from spatially resolved SED fitting and associated to thermal dust and radio emission in regions A to G (see  Fig.~\ref{fig:seds}).}
    \label{tab:spatial_sed}
    \begin{tabular}{lccccccc}
    \hline\hline
    \textbf{Parameters}   & \textbf{A} & \textbf{B} & \textbf{C} & \textbf{D} & \textbf{E} & \textbf{F} & \textbf{G} \\ \hline
    $\Sigma_\mathrm{dust}$~[M$_{\odot}$~pc$^{-2}$]        & 0.137~$\pm$~0.002  & 0.179~$\pm$~0.002  & 0.118~$\pm$~0.001  & 0.216~$\pm$~0.002  & 0.117~$\pm$~0.001  & 0.109~$\pm$~0.001  & 0.126~$\pm$~0.001 \\
    $\Sigma_\mathrm{small~grains}$~[M$_{\odot}$~pc$^{-2}$] & 0.0132~$\pm$~0.0002 & 0.0156~$\pm$~0.0002  & 0.0090~$\pm$~0.0001  & 0.0213~$\pm$~0.0003  & 0.0112~$\pm$~0.0002  & 0.0146~$\pm$~0.0002  & 0.0164~$\pm$~0.0002 \\
    $\Sigma_\mathrm{large~grains}$~[M$_{\odot}$~pc$^{-2}$] & 0.123~$\pm$~0.002  & 0.163~$\pm$~0.002  & 0.109~$\pm$~0.001  & 0.194~$\pm$~0.002  & 0.105~$\pm$~0.001  & 0.095~$\pm$~0.001  & 0.110~$\pm$~0.001  \\
    $\Sigma_\mathrm{gas}$/$\Sigma_\mathrm{dust}$   &  224.2~$\pm$~9.4  & 465.7~$\pm$~22.0  & 890.4~$\pm$~43.7  & 423.2~$\pm$~19.6  & 612.6~$\pm$~29.0  & 442.5~$\pm$~18.9  & 291.8~$\pm$~12.8   \\
    $T_\mathrm{dust}$~[K]        & 25.776~$\pm$~0.069  & 29.816~$\pm$~0.069  & 31.563~$\pm$~0.072  & 28.427~$\pm$~0.060  & 29.752~$\pm$~0.068  & 25.891~$\pm$~0.057  & 24.047~$\pm$~0.054   \\
    $\Sigma L_\mathrm{star}$~[$10^3$~L$_{\odot}$~pc$^{-2}$]        &  0.114~$\pm$~0.004  & 0.406~$\pm$~0.010  & 1.5659~$\pm$~0.025  & 0.357~$\pm$~0.009  & 0.845~$\pm$~0.015  & 0.177~$\pm$~0.004  & 0.180~$\pm$~0.003  \\
    $\Sigma L_\mathrm{dust}$~[$10^2$~L$_{\odot}$~pc$^{-2}$]        & 1.991~$\pm$~0.011  & 6.168~$\pm$~0.033  & 5.106~$\pm$~0.026  & 5.368~$\pm$~0.028  & 3.815~$\pm$~0.019  & 1.668~$\pm$~0.007  & 1.260~$\pm$~0.006 \\
    $\langle U\rangle$~[$2.2\times10^5$ Wm$^{-2}$]          & 7.267~$\pm$~0.113  & 16.883$\pm$~0.230  & 23.478~$\pm$~0.296  & 12.808~$\pm$~0.158  & 16.675~$\pm$~0.219  & 7.457~$\pm$~0.100  & 4.862~$\pm$~0.065 \\
    $q_\mathrm{AF}$          &  0.098~$\pm$~0.001  & 0.087~$\pm$~0.001  & 0.076~$\pm$~0.001  & 0.098~$\pm$~0.001  & 0.097~$\pm$~0.001  & 0.127~$\pm$~0.001  & 0.127~$\pm$~0.001 \\
    $\beta$         & 2.200~$\pm$~0.023  & 2.396~$\pm$~0.025  & 2.488~$\pm$~0.013  & 2.475~$\pm$~0.019  & 2.491~$\pm$~0.009  & 2.494~$\pm$~0.005  & 2.495~$\pm$~0.005 \\
    $\alpha_\mathrm{s}$           &  0.548~$\pm$~0.033  & 0.589~$\pm$~0.016  & 0.715~$\pm$~0.019  & 0.581~$\pm$~0.017  & 0.662~$\pm$~0.020  & 0.674~$\pm$~0.020  & 0.749~$\pm$~0.023 \\ 
    \hline
    \end{tabular}
\end{table*}

\bsp	
\label{lastpage}
\end{document}